\documentclass[aps,twocolumn,pre,amsmath,amssymb,notitlepage,superscriptaddress,longbibliography,nofootinbib,longbibliography]{revtex4-2}

\usepackage{graphicx}
\usepackage{dcolumn}
\usepackage{bm}

\usepackage[mathlines]{lineno}

\usepackage{color,xcolor}
\usepackage{amssymb}
\usepackage{amsmath}
\usepackage{mathtools}
\usepackage{tikz}
\usepackage{soul}

\usepackage{hyperref}
\hypersetup{
    colorlinks=true,
    linkcolor=blue,
    urlcolor=magenta,
    citecolor=red,
    pdfpagemode=FullScreen,
    }
\colorlet{shadecolor}{gray!20}

\newcommand{\cD}{\mathcal{D}}

\newcommand{\cI}{\mathcal{I}}

\newcommand{\cN}{\mathcal{N}}
\newcommand{\cP}{\mathcal{P}}

\newcommand{\cS}{\mathcal{S}}

\newcommand{\cO}{\mathcal{O}}

\newcommand{\bbeta}{\bar{\beta}}

\newcommand{\bgamma}{\bar{\gamma}}
\newcommand{\brho}{\bar{\rho}}

\newcommand{\bI}{\mathbf{I}}

\newcommand{\bfQ}{\mathbf{Q}}

\newcommand{\bfV}{\mathbf{V}}

\newcommand{\bfe}{\mathbf{e}}

\newcommand{\bfm}{\mathbf{m}}

\newcommand{\bfq}{\mathbf{q}}
\newcommand{\bfr}{\mathbf{r}}
\newcommand{\bfu}{\mathbf{u}}

\newcommand{\bfgamma}{\boldsymbol{\gamma}}
\newcommand{\bfbeta}{\boldsymbol{\beta}}
\newcommand{\bfrho}{\boldsymbol{\rho}}

\newcommand{\bfLL}{\mathbf{\Lambda}}
\newcommand\bone{{\bf \hat{1}}}

\newcommand{\bfhe}{\hat{\bfe}}

\newcommand\mfu{\mathfrak{u}}
\newcommand\rmd{\mathrm{d}}

\newcommand{\llangle}{\left\langle}
\newcommand{\rrangle}{\right\rangle}

\usepackage{fontawesome5}
\usepackage{wasysym}

\begin{document}

\title{Random motility regulation as a generic mechanism of community formation}

\author{Alberto Dinelli}
\affiliation{Department of Biochemistry, University of Geneva, 1211 Geneva, Switzerland}
\affiliation{Université Paris Cité, MSC, UMR 7057 CNRS, 75013 Paris, France}
\author{Ada Altieri}
\altaffiliation[Corresponding authors.]{ Email: {ada.altieri@u-paris.fr and jgt@mit.edu}}
\affiliation{Université Paris Cité, MSC, UMR 7057 CNRS, 75013 Paris, France}
\author{Julien Tailleur}
\altaffiliation[Corresponding authors.]{ Email: {ada.altieri@u-paris.fr and jgt@mit.edu}}
\affiliation{Department of Physics, Massachusetts Institute of Technology, Cambridge, Massachusetts 02139, USA}

\date{\today}

\begin{abstract} 
The self-organization of microbial ecosystems involves a large variety
of mechanisms, ranging from biochemical signaling to population
dynamics. Among these, the role of motility regulation has been little
studied, despite the importance of active migration processes.  
Here
we show how weak, random motility regulation suffices to induce complex forms of organization in bacterial mixtures comprising a large number of
coexisting strains. 
First, we simulate microscopic models of
run-and-tumble bacteria whose self-propulsion speeds are weakly regulated by
the local density of each strain, mimicking the impact of weak, random metabolic interactions. Our simulations reveal that, as the
heterogeneity of the interaction network increases, the system
undergoes a phase transition leading to the emergence of distinct, spatially segregated
communities. To account for these results and assess their robustness,
we use random-matrix theory to analyze the hydrodynamic description of
the bacterial mixture, obtaining a quantitative agreement with our
microscopic simulations. Our results hold for a variety of motility-regulation mechanisms and highlight the need to characterize the role of motility regulation in experimentally relevant situations.
\end{abstract}

\maketitle

\tableofcontents

\section{Introduction}
\noindent Nature around us is not invariant by translation but instead consists of a variety of spatially segregated ecosystems. 
How such a form of spatial organization emerges is an open problem in ecology that involves a variety of mechanisms as diverse as speciation~\cite{kopp2010speciation, hubbell2011}, migration~\cite{allesina2020going, leibold2004, wilcove2008going}, interaction with the environment~\cite{gaston2000, tsai2014, dethlefsen2008, ratzke2018modifying,daniels2023}, or competition for resources~\cite{hardin1960competitive, grime1973competitive, ho2024resource, barabas2016effect}. 
From a physics perspective, this raises the question of the phase transitions governing the self-organization of proliferating active matter~\cite{hallatschek2023proliferating}.
In this context, microbial ecosystems have received a lot of attention, both because of their abundance in nature and because they offer a controlled setting to study community formation~\cite{friedman2017community,good2017dynamics,pearce2020stabilization,hu2022emergent,ho2024resource,hatton2024diversity}. 

A lot of progress has been made in the study of low-dimensional ecosystems, in which a small number of strains coexist~\cite{vano2006, vet2018bistability, blanchard2015, van2015_competition}. Much focus has been brought on how genetic drift, by which the random extinction of bacterial strains impacts the local composition of the colony, induces genetic sectoring and community formation~\cite{hallatschek_genetic_2007,korolev2010genetic}. 
In addition, competition for resources or mutualistic interactions that impact growth rates have also been extensively studied since the work of Lotka and Volterra~\cite{hofbauer1998,lotka2013}. 

However, it is also crucial to study more realistic, high-dimensional ecosystems, which comprise a large number of coexisting species~\cite{may1972will,bunin2017ecological,tikhonov2017,galla2018dynamically,barbier_generic_2018,hu2022emergent}. 
This is an important challenge since it is unclear when lessons from the low-dimensional limit generalize to complex ecosystems~\cite{bloxham2022diauxic,bloxham2024}.
For instance, studies have shown that high-dimensional generalizations of the Lotka-Volterra and Resource-Consumer models display complex attractors that cannot be extrapolated from low-dimensional systems~\cite{tikhonov2017, bunin2017ecological, biroli2018marginally, altieri2021properties, altieri2022effects, cui2020, blumenthal2024,salvatore2025patterns}, suggesting that, in ecology also, ``more is different"~\cite{anderson1972more}.

While there are many established results on how the regulation of growth and death can induce community formation, 
far less is known about the mechanisms relying on motility regulation. 
This is partly explained by the experimental difficulty of quantifying motility at the microscopic scale~\cite{figueroa20203d,kurzthaler2024characterization}. 
This is an important gap since motility regulation, in the form of chemotactic~\cite{budrene_complex_1991} or quorum-sensing~\cite{miller2001QS,liu_sequential_2011,liu2013phase,curatolo_cooperative_2020} interactions, has been established as a powerful pattern-formation pathway in low-dimensional ecosystems.

On the contrary, the role played by motility regulation remains unchartered territory in high-dimensional ecosystems with many co-existing species. There, in addition to specific quorum-sensing signaling pathways, motility regulation also emerges as a byproduct of metabolic interactions, whereby bacteria exchange metabolites such as vitamins, amino acids, nucleotides, and growth factors~\cite{morris2013syntrophy,d2018ecology}. 
Motility regulation then results from the modulation of the available energy: \textit{E.~coli} uses around 5\% of the metabolic output for propulsion~\cite{schavemaker2022flagellar}. It may also stem from more elaborate adaptation mechanisms of bacteria swimming gaits to metabolic indicators through protein sensing or energy taxis~\cite{alexandre2004ecological,seymour2010chemotactic,porter2011signal,alvarado2020protein,keegstra2022ecological}.

\begin{figure}
    \begin{center}
      \includegraphics[width=\columnwidth]{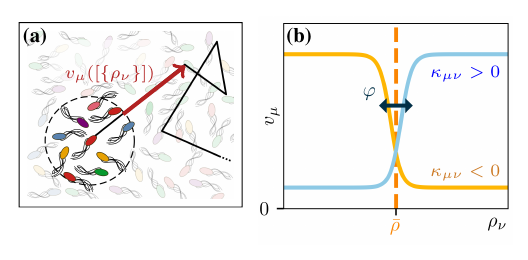}
    \end{center}
    \caption{{Sketch of the run-and-tumble dynamics and of the motility regulation mechanism.} ({\bf a}) Sketch of the dynamics. The different colors correspond to the $N$  different strains and the speed $v_\mu$ of strain $\mu$ depends on the local density of bacteria in each strain. ({\bf b}) Motility regulation of strain $\mu$ by strain $\nu$ through Eq.~\eqref{eq:v_mu_random} in a two-strain system. }
  \label{fig:sketch}
\end{figure}

In this article, we study the impact of the random motility regulation that results from such complex interaction networks on spatially extended ecosystems. 
We show that even a very weak random motility regulation suffices to drive the phase separation of high-dimensional ecosystems, resulting in a diverse set of coexisting communities. 
Remarkably, this fragmentation of the ecosystem into distinct phases happens even in the presence of pre-existing attractive interactions between the strains: A gregarious community---where all strains coexist--- also becomes fragmented upon adding weak random motility regulation. 

To demonstrate our results, we introduce below a  model accounting for random motility regulation between $N\gg 1$ strains of run-and-tumble bacteria~\cite{berg_e_2004}, which we study numerically in two spatial dimensions. 
Namely, the self-propulsion speed of strain $\mu$ is chosen to depend on the local density of all strains, $v_\mu[\{\rho_\nu\}]$ (Fig.~\ref{fig:sketch}a-b), in such a way that $\{\partial_{\rho_\nu} v_\mu\}$ forms a set of random variables. 
Our main result is that the system undergoes a fragmentation transition as the variance of the random interaction increases and that this happens for a small variance $\sim \sigma^2/N$ (Fig.~\ref{fig:snapshots}).
To account for this transition, we coarse-grain our microscopic model and employ random-matrix theory to characterize the phase behavior predicted by the resulting hydrodynamic theory, in the spirit of~\cite{sear2003instabilities,shrinivas2021phase,thewes2023composition,parkavousi2025enhanced}. We then validate our results using extensive numerical simulations in one space dimension. Finally, we show the robustness of our results by considering different motility-regulation mechanisms, including both reciprocal and non-reciprocal interactions, as well as different topologies of the interaction network. Our work thus establishes random motility regulation---a form of interaction which we argue should be generic in microbial ecosystems due to metabolic interactions~\cite{morris2013syntrophy,d2018ecology,schavemaker2022flagellar,alexandre2004ecological,seymour2010chemotactic,porter2011signal,alvarado2020protein,keegstra2022ecological}---as a generic mechanism leading to the spatial self-organization of complex bacterial mixtures, even when this regulation is very weak. 
Despite the intrinsic difficulty of characterizing motility regulation~\cite{figueroa20203d,kurzthaler2024characterization}, our results thus establish that this is an important experimental goal.

\section{Random-motility-regulation model}
\label{sec:model-RBM}
We consider $N$ strains of bacteria, whose motions alternate between 
runs and tumbles. During a run phase, bacteria $i$ of strain $\mu$ self-propels with fixed orientation 
$\bfu_{i,\mu}$ so that its position evolves as
$\dot{\bfr}_{i,\mu} = v_{i,\mu} \bfu_{i,\mu}$. Tumble events correspond to full randomizations of the bacteria's orientations, and occur with a constant rate $\tau^{-1}$, as sketched in Fig.~\ref{fig:sketch}a.
To crudely account for the motility regulation induced by metabolic interactions, 
we associate  a `metabolic state' $\cI_{i,\mu}$ with each bacterium $(i,\mu)$, which models the energy available for the bacteria's biological functions, including motility. 
In a well mixed configuration where all strains have density $\rho_\mu=\bar \rho$, we denote the metabolic state as $\cI_{\mu}^0$. 
Within linear response, we predict density fluctuation to impact the metabolic state through:
\begin{equation}
    \cI_{i,\mu}(\bfr_{i,\mu}, [\{\rho_\nu\}]) = \cI^0_\mu + \sum_{\nu=1}^N \kappa_{\mu\nu} [\tilde \rho_\nu(\bfr_i)-\bar\rho]\;,
    \label{eq:metabolic_level}
\end{equation}
where the density is measured locally as $\tilde \rho_\nu\equiv K\ast \rho_\nu$, with $K$ a convolution kernel representing the transport of signaling molecules or metabolites. The matrix $\kappa_{\mu\nu}$ thus couples the metabolic state of strain $\mu$ to density fluctuations of strain $\nu$ and we use the range $\ell_{\rm int}$ of $K$ as the unit of length. Equation~\eqref{eq:metabolic_level} assumes that metabolic interactions are fast, allowing us to eliminate the mediating chemical fields, as detailed in Appendix~\ref{app:motility_regulation}. Broader choices of interactions are possible~\cite{smith2003animal,colombo2023pulsed} and we expect the mechanism at the core of our results to extend beyond the cases discusses here. 

In turns, the motility of bacterium $(i,\mu)$ depends on its current metabolic state. We assume that each bacterium moves with a self-propulsion speed $v_0^\mu$ in the reference state, and model motility regulation as a non-linear response to fluctuations of the metabolic state:
\begin{equation}
    v_{i,\mu} = {v}_{\mu} \left(\Delta \cI_{i,\mu}\right) \;, \qquad \Delta \cI_{i,\mu} \equiv  \cI_{i,\mu} -  \cI_{\mu}^0 \;,
  \label{eq:v_mu_metabolic}
\end{equation}
where ${v}_{\mu}(x)$ is a positive function satisfying ${v}_{\mu}(0)=v_0^\mu$. For the sake of simplicity,  we set $v_0^\mu=v_0$ for all strains in the following. Specifically, we choose:
\begin{equation}
  v_\mu(\bfr_{i,\mu}, [\{\rho_\nu\}]) = v_0 \cS \left(\sum_{\nu=1}^N \kappa_{\mu\nu} \frac{\tilde \rho_\nu(\bfr_i)-\bar{\rho}}{\varphi}\right) \;,
  \label{eq:v_mu_random}
\end{equation}
where $\cS$ is a sigmoidal function and $\varphi$ controls the sensitivity of motility to the variations of the energy available to the bacterium. (See Appendix~\ref{app:motility_regulation} for a discussion of Eq.~\eqref{eq:v_mu_random}.)
All in all, metabolic interactions are modelled through the effective density-based motility regulation given by Eq.~\eqref{eq:v_mu_random}:  Strain $\nu$ then either enhances ($\kappa_{\mu\nu} > 0$) or inhibits ($\kappa_{\mu\nu} < 0$) the self-propulsion speed of strain $\mu$ as depicted schematically in Fig.~\ref{fig:sketch}b. We note that our results are robust to the precise choice for the functional form of Eq.~\eqref{eq:v_mu_random}, as discussed in Section~\ref{sec:extension}.

In most of this article, we model the random motility-regulation using independent Gaussian-distributed
$\kappa_{\mu\nu}$. This choice allows for analytical progress, and, as often with random-matrix theory, we expect that our results extend to  distributions with finite means and variances~\cite{biroli2018marginally}. The distribution of $\kappa_{\mu \nu}$ is then entirely determined by its first two moments, which we define as:
\begin{equation}
  \langle \kappa_{\mu \nu} \rangle = \dfrac{m}{N} \;, \qquad
  \text{Var}[\kappa_{\mu \nu}] = \frac{\sigma^2}{N}\;.
  \label{eq:kappa_munu}
\end{equation}
We first note that the scaling of $\langle \kappa_{\mu \nu} \rangle$ simply ensures that the sum in Eq.~\eqref{eq:v_mu_random} remains finite in the large-$N$ limit. The scaling of the variance will prove to be such that the fragmentation transition occurs for $\sigma^2 \sim \cO(1)$: \textit{Weak} random motility regulation thus suffices to drive community formation. Different scalings can then be extrapolated by sending $\sigma^2$ to $0$ or to $\infty$, as discussed in the Appendix~\ref{app:scaling}. Below, we focus on the case $m\leq 0$, where most of the interesting physics happens, which is also the most biologically relevant case since we expect competition to make the self-propulsion of species $\mu$ decrease on average when the densities of the other species increase~\cite{liu2019self}. 

\section{Numerical simulations in 2d}
\label{sec:2dsimul}

\begin{figure}
    \begin{center}
      \includegraphics[width=\columnwidth]{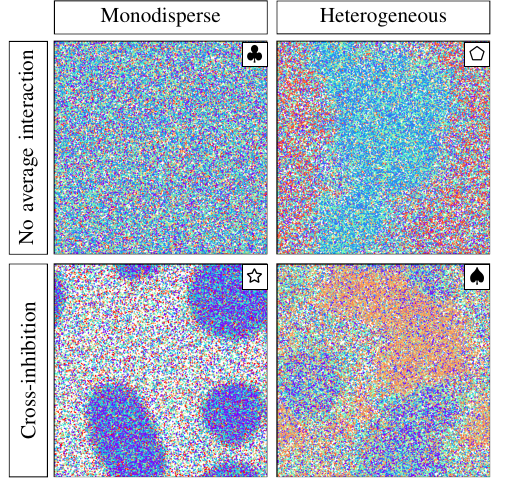}
    \end{center}
    \caption{{Snapshots from $2d$ microscopic simulations with $N=50$ strains for a random realization of $\{\kappa_{\mu\nu}\}$}. Bacteria of the same strain have the same color. For clarity, we show $10$ strains out of 50. Top panels ($\clubsuit$, $\pentagon$): Phase behavior in the absence of average motility regulation, $m=0$. ($\clubsuit$): In the noninteracting limit $\sigma^2=0$, the system is homogeneous. ($\pentagon$): A fragmented ecosystem emerges upon increasing the heterogeneity ($\sigma^2 = 2$). Bottom panels ({\scriptsize \faStar[regular]},$\spadesuit$): Simulations with average motility inhibition $m = -1$. ({\scriptsize \faStar[regular]}): We observe gregarious phase separation in the monodisperse limit $(\sigma^2 = 0)$. ($\spadesuit$): Increasing heterogeneity again leads to fragmentation and community formation ($\sigma^2=2$). Parameters common to all
    panels: $L_x = L_y = 20$, average density per species $\rho_0 = 15$, $\bar{\rho} = 15, \varphi=10$, $v_0=1$, $\tau=1$. Additional details can be found in Appendix~\ref{app:heatmps}.}
  \label{fig:snapshots}
\end{figure}
To understand how random motility regulation impacts the large-scale
behavior of the system, we start by performing $2d$ simulations of the
bacterial dynamics described above (see App.~\ref{app:simulations},~\ref{app:heatmps} for details). In simulations, we vary the average inter-strain motility inhibition $m$ and the interaction heterogeneity
$\sigma^2$. As we show in Fig.~\ref{fig:snapshots}, a rich variety of phases emerges from random motility regulation.

We first show results for an ecosystem without systematic
regulation ($m=0$) in the top panels of
Fig.~\ref{fig:snapshots}. In the monodisperse case ($\sigma^2=0$), the
homogeneous phase is stable. Upon increasing $\sigma^2$, we observe a transition into a fragmented ecosystem: The steady state is phase-separated and comprises coexisting communities that differ in strain compositions {(see Movie S1)}.

To test whether random motility regulation can compete with a pre-existing self-organization of the colony, we consider the case of a systematic motility
inhibition ($m < 0$) in the bottom panels of
Fig.~\ref{fig:snapshots}. In the monodisperse
case, a gregarious phase is observed when $m$ is sufficiently negative:
All species then equally participate to a motility-induced phase separation~\cite{cates_motility-induced_2015,curatolo_cooperative_2020}{, as shown in Movie S2}. Upon
increasing $\sigma^2$, the colony transitions again into a fragmented
ecosystem, hence overcoming the aggregation induced by
$m<0$ {(see Movie S3)}. A weak, random motility regulation is thus sufficient to drive community
formation in microbial ecosystems, both in the absence and in the
presence of a pre-existing self-organization. Note that, for a given
choice of $(m, \sigma^2)$, the emergent behavior described above is robustly
observed across simulations with different realizations of the random coefficients $\{\kappa_{\mu\nu}\}$.

\begin{figure}
    \begin{center}
      \includegraphics[width=\columnwidth]{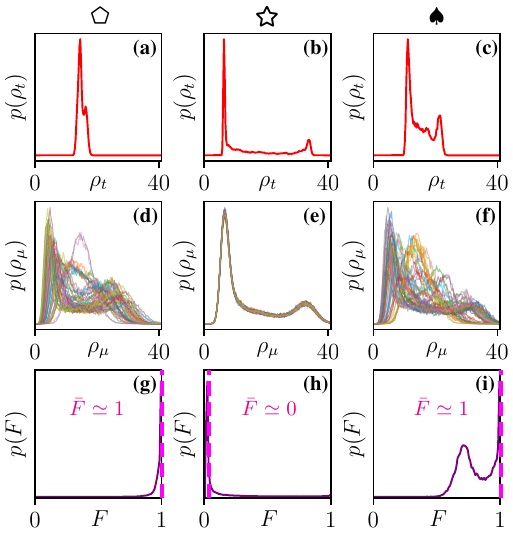}
    \end{center}
    \caption{{Statistical analysis performed on simulations from Fig.~\ref{fig:snapshots}.} Pentagon ($\sigma^2 = 2$, $m = 0$), star~($\sigma^2 = 0$, $m = -1$) and clubsuit ($\sigma^2 = 2$, $m = -1$) symbols in each column match the corresponding simulations in Fig.~\ref{fig:snapshots}.
    (a-c): Histograms of the total density $\rho_{\rm t}$. The overall modulation of the ecosystem density stems only from the average motility inhibition (b) and recedes upon increasing heterogeneity (a, c). 
    (d-f): Each single-strain density histogram is shown using a different color; Heterogeneity suffices to induce single-strain phase separation even in the absence of average inhibition (d).
    (g-i): Histograms of the fragmentation parameter $F$ and most probable value $\bar F$, which show the transition to fragmentation as heterogeneity increases. Additional details can be found in Appendix~\ref{app:heatmps}.}
  \label{fig:histograms}
\end{figure}
To characterize the transition to fragmented ecosystems, 
we first measure the histograms of the local density fields $\rho_\mu(\bfr)$, as well as that of the total density $\rho_{\rm t} \equiv N^{-1} \sum_\nu \rho_\nu$. The corresponding measurements are shown in Fig.~\ref{fig:histograms}. When random motility regulation is introduced in a non-interacting system, Fig.~\ref{fig:histograms}{$\pentagon$}, the emergence of the fragmented phase is reflected in the bimodal nature of the single-strain distributions $p(\rho_{\nu})$, Fig.~\ref{fig:histograms}d. The total density, however, remains essentially homogeneous and $p(\rho_{\rm t})$ is barely altered, Fig.~\ref{fig:histograms}a. The fragmentation transition then resembles a demixing transition. On the contrary, the gregarious phase observed in a monodisperse system, Fig.~\ref{fig:histograms}{\faStar[regular]}, leads to two peaks in the histogram $p(\rho_{\rm t})$, Fig.~\ref{fig:histograms}b, identical to those observed at the single-species level, Fig.~\ref{fig:histograms}e,. Finally, when heterogeneity is introduced in a gregarious system, Fig.~\ref{fig:histograms}{$\spadesuit$}, one  observes both a reminiscence of the global phase separation in $p(\rho_{\rm t})$, Fig.~\ref{fig:histograms}c, and a rich variety of distributions at the single-strain level $p(\rho_{\nu})$, Fig.~\ref{fig:histograms}f,: the system experiences a combination of demixing and phase separation.

To better characterize the differences between the gregarious and fragmented ecosystems, we  introduce a fragmentation order parameter. To do so, we  consider local density
fluctuations around the mean density $\rho_0$ of each strain as
$\mathbf{\delta}\bfrho(\bfr)=(\delta \rho_1, \dots, \delta\rho_N)$, where $\delta\rho_\mu(\bfr) \equiv \rho_\mu(\bfr)-\rho_0$. In the perfectly gregarious case at $\sigma^2=0$, all fluctuations $\delta\rho_\mu(\bfr)$ occur synchronously in space, so that  $\delta\bfrho(\bfr)$ is parallel 
to $\bone = (1,\dots,1)/\sqrt{N}$. On the contrary, in the fragmented case, density fluctuations of non-coexisting strains have opposite signs.
To distinguish these two scenarios, we thus use the relative angle between $\mathbf{\delta}\bfrho(\bfr)$ and $\bone$, as was done to characterize equilibrium phase-separation in multi-component systems~\cite{sear2003instabilities,shrinivas2021phase}. This leads us to define a fragmentation order-parameter field as:
\begin{equation}
  F(\bfr) \equiv 1 - \cos^2 \theta(\bfr) \;, \quad
  \cos^2\theta(\bfr) = \frac{|\delta \bfrho(\bfr) \cdot
    \bone|^2}{|\delta\bfrho(\bfr)|^2} \;.
  \label{eq:F_parameter}
\end{equation}
When $F(\bfr) \approx 1$, the system is thus locally fragmented, while
$F(\bfr) \approx 0$ corresponds to a locally gregarious behavior.

We
report the histogram of $F(\bfr)$ in the bottom
panels of Fig.~\ref{fig:histograms}. As expected, $p(F)$ is
peaked close to $0$ in the gregarious phase (h), whereas it is peaked close to $1$
in the fragmented case (g). In panel (i), the double-peak structure stems from a spatial coexistence between fully demixed regions and partially gregarious communities that are becoming more fragmented as $\sigma^2$ increases. The spatial structures corresponding to these histograms are illustrated in Appendix~\ref{app:heatmps}. All in all, the fragmentation
parameter $F$ is thus a relevant order parameter to distinguish gregarious and fragmented communities.

Our simulations thus demonstrate the existence of a phase transition
towards a fragmented ecosystem upon increasing the heterogeneity of
random motility-regulation interactions. We now develop a theoretical framework to elucidate the physical origin of this transition, starting from the bacteria microscopic dynamics.

\section{Theory}\label{sec:theory}
Standard methods~\cite{dean_langevin_1996,tailleur_statistical_2008,cates_when_2013,solon_2018_generalized,curatolo_cooperative_2020,dinelli2023nonreciprocity,duan2023dynamical,dinelli2024fluctuating} recapitulated in Appendix~\ref{app:coarsegraining} show that, at the macroscopic scale, the run-and-tumble dynamics lead to a diffusive behavior and the evolution of the density field of strain $\mu$ reads
\begin{equation}
    \partial_t \rho_\mu = \nabla_{\bfr} \cdot \left[ M_\mu \nabla_{\bfr} \mfu_\mu \right]\;, 
    \label{eq:hydro}
\end{equation}
where $M_\mu=\rho_u v_\mu^2\tau/d$ is the collective mobility of strain $\mu$, $d$ the number of space dimensions, and
$\mfu_\mu=\log v_\mu(\{\rho_\nu(\bfr)\}) + \log \rho_\mu(\bfr)$ plays the
role of an effective chemical potential. The entropic contribution $\log \rho_\mu$ accounts for the random-walk nature of run-and-tumble dynamics that favors homogeneous steady states. On the contrary, the excess chemical potential
$\log v_\mu(\bfr,[\{\rho_\nu\}])$ stems from the motility regulation and reflects the tendency of bacteria to accumulate where they move slower. Note that the distinct
fields are only coupled through the density dependence of their motilities.
In the absence of motility regulation, Eq.~\eqref{eq:hydro} describes $N$ non-interacting, diffusive fields. Even if the field diffusivities are distributed, the uniform state is always stable.

To predict the instability of homogeneous states, we first note that, when the variations of the density field occurs on scales much larger than the range of the kernel $K$, we can use a local approximation: $\tilde\rho_\mu(\bfr) = \rho_\mu(\bfr)$. Physically, this means that the bacterial density varies on scales larger than the length over which chemical signals are transported before they are too dilute to be detected. Mathematically, it allows us to write the self-propulsion speed as a function, not a functional, of the densities: $v_\mu(\bfr)=v_\mu(\{\rho_\nu(\bfr)\}])$. We then perform a linear-stability analysis of the corresponding `local' field theory around a well-mixed homogeneous state: $\bfrho_0 \equiv \rho_0
(1, \dots, 1)$. In Fourier space, the linearized dynamics of a density
fluctuation reads: $\partial_t
\delta\hat\rho_{\mu}(\bfq) = -q^2 \mathcal{D}_{\mu\nu}
\delta\hat\rho_{\nu}(\bfq)$, where $\bfq$ denotes the Fourier mode and
$\cD_{\mu\nu}$ is a diffusivity matrix given by:
\begin{equation}
  \cD_{\mu\nu}=\cD_\mu^0\left[\delta_{\mu\nu}+ \rho_0
    \frac{\partial\log v_\mu}{\partial \rho_\nu} \right] , \;\;\cD_{\mu}^0 = \frac{v_{\mu}(\bfrho_0)^2 \tau}{d}\;.
  \label{eq:general_LSA}
\end{equation}
In the absence of
motility-regulation, Eq.~\eqref{eq:general_LSA} amounts to $N$ independent diffusion equations with diffusivity $\cD_{\mu}^0$. Motility
regulation couples the dynamics of all strains through the terms
$\partial_{\rho_\nu} \log v_\mu$. A homogeneous well-mixed state becomes
unstable when the eigenvalue of $\cD_{\mu\nu}$ with the smallest real
part, denoted by $\lambda_0$, satisfies Re$(\lambda_0)<0$. The corresponding normalized eigenvector $\bfhe_0$ can then be used to estimate the 
fragmentation parameter as
\begin{equation}
    F(\bfhe_0)=1-|\bfhe_0 \cdot \bone|^2\;.
    \label{eq:F_eigenv}
\end{equation}

\begin{figure*}
    \begin{center}
      \includegraphics[width=\textwidth]{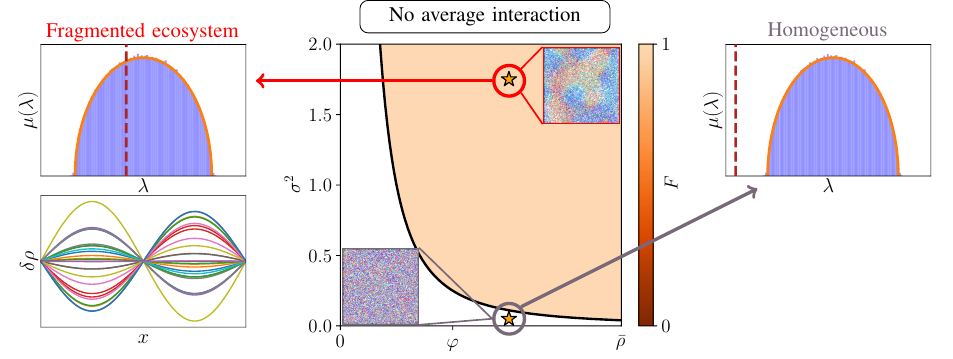}
    \end{center}
    \caption{{Phase diagram in the $(\rho_0,\sigma^2)$-space for symmetric interactions $\kappa_{\mu\nu}=\kappa_{\nu\mu}$ and $\rho_0= \brho$: no average interaction ($m = 0$).} White regions represent stable homogeneous phases, while shaded regions correspond to phase-separated states. The color encodes the value of the fragmentation order parameter $F$ predicted by Eq.~\eqref{eq:F_eigenv}, which ranges from $0$ in gregarious ecosystems to $1$ in fragmented ones. For $m=0$, Eq.~\eqref{eq:transition1} (black line) predicts a transition from homogeneous to fragmented ecosystems for large enough $\sigma^2$. Two illustrative data points are plotted with star symbols. For each, we show in inset a steady-state snapshot of the system, obtained from simulation in a $16\times16$ domain. For clarity, we report $25$ out of $N=100$ strains in each snapshot. We then also plot the spectral densities $\mu(\lambda)$ predicted theoretically (orange line) and computed numerically (lilac histogram). The dashed red line indicates the instability threshold $\lambda=0$. When the system is predicted to be unstable, we  plot the density modulation corresponding to the most unstable eigenvector~$\bfhe_0$. The associated eigenvector~$\bfhe_0$ is orthogonal to $(1, \dots, 1)$, corresponding to a spatial demixing of the strains. }
 \label{fig:phasediag_m0}
\end{figure*}

\begin{figure*}
    \begin{center}
      \includegraphics[width=\textwidth]{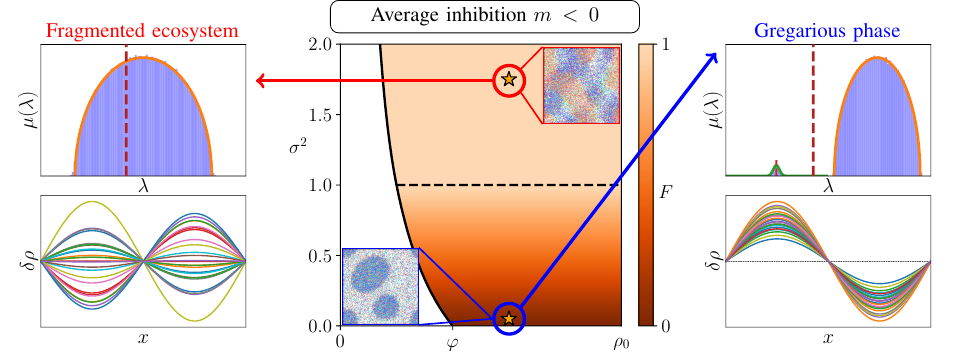}
    \end{center}
    \caption{{Phase diagram in the $(\rho_0,\sigma^2)$-space for symmetric interactions $\kappa_{\mu\nu}=\kappa_{\nu\mu}$ and $\rho_0= \brho$: average inhibition ($m < 0$).} White regions represent stable homogeneous phases, while shaded regions correspond to phase-separated states. The color encodes the value of the fragmentation order parameter $F$ predicted by Eq.~\eqref{eq:F_eigenv}, which ranges from $0$ in gregarious ecosystems to $1$ in fragmented ones. For~$m < 0$, the system is first gregarious at small $\sigma^2$ and becomes increasingly fragmented as the heterogeneity increases towards the BBP transition line at $\sigma^2=m^2$ where $F=1$ (dashed line). The boundaries of the homogeneous phase (solid lines) are predicted by Eqs.~\eqref{eq:transition1} and~\eqref{eq:transition2} for $\sigma^2>m^2$ and $\sigma^2<m^2$, respectively. Two illustrative data points are plotted with star symbols. For each, we show in inset a steady-state snapshot of the system, obtained from simulation in a $16\times16$ domain. For clarity, we report $25$ out of $N=100$ strains in each snapshot. We then also plot the spectral densities $\mu(\lambda)$ predicted theoretically (orange and green lines) and computed numerically (lilac histogram). The dashed red line indicates the instability threshold $\lambda=0$. When the system is predicted to be unstable, we  plot the density modulation corresponding to the most unstable eigenvector~$\bfhe_0$. For $\sigma^2<m^2$, the spectral density includes an outlier located to the left of the Wigner semicircle (top right inset), which first becomes unstable. At $\sigma^2=0$, the associated eigenvector is $\bfhe_0 \propto (1, \dots, 1)$, leading to a gregarious dynamics.  For $\sigma^2>m^2$, the Wigner semicircle has swallowed the outlier and $\bfhe_0\cdot \bone=0$. }
 \label{fig:phasediag_mfinite}
\end{figure*}

\begin{figure}
    \begin{center}
      \includegraphics[width=\columnwidth]{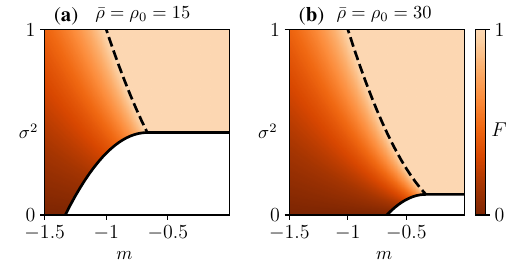}
    \end{center}
    \caption{Analytical phase diagram in the $(m,\sigma^2)$-space for symmetric interactions $\kappa_{\mu\nu} = \kappa_{\nu\mu}$ and $\rho_0=\bar\rho$. White regions represent stable homogeneous phases, while shaded regions correspond to phase-separated states. In both panels, the color encodes the value of the fragmentation order parameter $F$ predicted by Eq.~\eqref{eq:F_eigenv}, which ranges from $0$ in gregarious ecosystems to $1$ in fragmented ones. Parameters: $\varphi=20$, $\rho_0=\brho$.}
    \label{fig:sigma_m_phasediag}
\end{figure}

To make progress analytically, we consider the case $\rho_0 = \brho$, which leads to a diffusivity matrix
$\cD_{\mu\nu}$ given by:
\begin{equation}
\cD_{\mu\nu} = \cD_0 \left(\delta_{\mu\nu} + \frac{\rho_0}{\varphi}
\kappa_{\mu\nu} \right) \;,
\label{eq:Diff_mat}
\end{equation}
where $\cD_0=v_0^2\tau/d$, see App.~\ref{app:linearized_field_theory}. Equation~(\ref{eq:Diff_mat}) then allows us to characterize analytically the spectrum of $\cD_{\mu\nu}$ using results from random-matrix theory~\cite{edwards1976eigenvalue,baik2005phase,paul2007asymptotics,benaych2011eigenvalues}. 
Furthermore, we first focus on the case
of symmetric interactions $\kappa_{\mu\nu}=\kappa_{\nu\mu}$, which make the eigenvalues $\lambda_i$ of $\cD$ real. 
We note that
these specific choices do not restrict the scope of our results: In
App.~\ref{app:numerical_phdiag} we extend our analysis to $\rho_0 \neq
\brho$. Let us now show how, starting from Eq.~\eqref{eq:Diff_mat}, we can predict the phase diagrams presented in Figs.~\ref{fig:phasediag_m0}--\ref{fig:sigma_m_phasediag}.

\subsection{The fragmentation transition.}
\label{sec:fragmentation}
To study the onset of the fragmentation transition, we consider the spectral density of eigenvalues $\mu(\lambda)
\equiv \lim \limits_{N \to \infty} \frac{1}{N} \sum_{i=1}^N \langle
\delta(\lambda-\lambda_i) \rangle$, where $\langle \cdot \rangle$
denotes the average over the realizations of $\kappa_{\mu\nu}$. We can use Eq.~\eqref{eq:Diff_mat} to infer the spectral density of the diffusivity matrix from that of the GOE matrix $\kappa_{\mu\nu}$~\cite{livan2018introduction}.

In the absence of pre-existing self-organization ($m=0$), the result is a shifted Wigner semicircle \cite{wigner1958distribution}, centered at $\lambda=\mathcal{D}_0$, and of radius $2\mathcal{D}_0\rho_0\sigma/\varphi$. The homogeneous phase is then predicted to be linearly unstable whenever the heterogeneity $\sigma^2$ is large enough:
\begin{equation}
    \sigma^2>\frac{\varphi^2}{4 \rho_0^2}\;.
    \label{eq:transition1}
\end{equation}
Furthermore, random-matrix theory tells us that the eigenvectors of $\cD_{\mu\nu}$ are uniformly distributed on the unit sphere. 
In the limit of large $N$, the eigenvectors are thus typically orthogonal to $\bone$ and the perturbations $\{\delta \rho_\mu\}$ are such that $\sum_\mu \delta\rho_\mu \simeq 0$: 
The system undergoes demixing between communities with different strain compositions and the fragmentation parameter is $F\simeq 1$. We report in Fig.~\ref{fig:phasediag_m0} the phase diagram predicted by the transition line given by Eq.~\eqref{eq:transition1}, together with
illustrative plots for the spectral density and the eigenvector components.

When an average self-inhibition is present ($m<0$), the spectrum of $\cD_{\mu\nu}$ comprises both the shifted Wigner semicircle and a Gaussian-distributed outlier centered at
$\lambda_{\rm 0} = \cD_0 \left[1 +  (m_{\rm} + \sigma^2/{m})\rho_0/\varphi \right]$, as
shown in the left inset of Fig.~\ref{fig:phasediag_mfinite}. As $N\to\infty$,  the outlier becomes delta-distributed. There are then two different scenarios, separated by the so-called BBP transition~\cite{edwards1976eigenvalue,baik2005phase}. For small heterogeneity, $\sigma^2<m^2$, the outlier is outside of the Wigner semi-circle, and is the smallest eigenvalue. The homogeneous phase is then unstable whenever:
\begin{equation}
    \rho_0 \sigma^2 > -m \varphi - m^2 \rho_0\;.
    \label{eq:transition2}
\end{equation}
The eigenvector associated to the outlier has been characterized~\cite{benaych2011eigenvalues} and satisfies $\bfhe_0 \cdot \bone=\sigma/|m|$ so that $F=\sigma^2/m^2$. For $\sigma^2=0$ we recover a gregarious phase with $F=0$, while the ecosystem becomes increasingly fragmented as $\sigma^2$ increases. At $\sigma^2=m^2$, the Wigner semicircle swallows the outlier. The physics of the system is then  entirely determined by the weak random-motility regulation through 
 the heterogeneity $\sigma^2$. The fragmentation transition occurs again when Eq.~\eqref{eq:transition1} is satisfied and leads to $F=1$. The transition lines corresponding to these two scenarios, together with illustrative spectral densities and eigenvectors, are shown in Fig.~\ref{fig:phasediag_mfinite} (solid black lines), along with the value of $\sigma$ at which the BBP transition occurs (dashed line). 
 The colormap for the
fragmentation parameter $F$ highlights how, upon increasing the
heterogeneity $\sigma^2$, fragmentation overrules any pre-existing
form of organization. Two phase diagrams in the $(m,\sigma^2)$ plane at fixed $\rho_0$ shows in Fig.~\ref{fig:sigma_m_phasediag} how the two scenarios above are connected as $m$ increases.

Random-matrix theory thus predicts how weak random motility regulation generically leads to the formation of fragmented ecosystem both in the presence and in the absence of pre-existing self-organization. These results are in qualitative agreement with the 2d numerical results shown in Fig.~\ref{fig:snapshots} and~\ref{fig:histograms}. We note that these predictions hold in the $N\to \infty$ limit, which is hard to statistically characterize in 2d. We thus now present results of extensive numerical simulations in 1d that allow testing more precisely the transition lines.

\begin{figure}
  \begin{center}
      \includegraphics[width=\columnwidth]{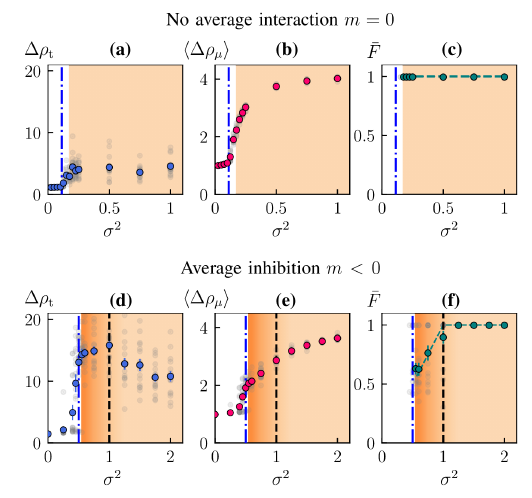}
  \end{center}
  \caption{{Statistical analysis of simulations in $d=1$ with
      symmetric interactions.} Simulations are performed at fixed $\rho_0=\brho$ for different values of the heterogeneity $\sigma^2$, both in the absence (\textbf{a-c}) and in the presence (\textbf{d-f}) of an average motility inhibition. For each value of $\sigma^2$
      the results of 15 simulations are shown as gray points, whose averages are shown using colored disks. White regions correspond to homogeneous simulations. Shaded regions correspond to fragmented or gregarious ecosystems and their colors encode the most probably fragmentation order parameter extracted from the simulations, using the same colorcode as in Figs.~\ref{fig:phasediag_m0},~\ref{fig:phasediag_mfinite}. Vertical dot-dashed blue lines are the theoretical predictions given in Eqs.~\eqref{eq:transition1} (top) and~\eqref{eq:transition2} (bottom) for the transitions. The vertical black dashed line is the BBP transition $\sigma^2=m^2$. The green dashed line is the prediction for the fragmentation order parameter, $F=1$ in fragmented ecosystems and $F=\sigma^2/m^2$ below the BBP transition.
      Symbols measure the normalized standard deviations of the total-density histogram (blue, panels a\&{}d) and of the single-strain density histograms (red, panels b\&{}e). The data are normalized by the Poissonian fluctuations observed in homogeneous states. The green symbols show the most probable values of the fragmentation order parameter $\bar  F$ as defined in Fig.~\ref{fig:histograms}. Parameters of the simulation: $N=100$, $L_x=30$, $\rho_0 = \brho = 30$, $v_0=1$, $\tau=1$. In (a), $m=0, \varphi=20$, while in (b) $m=-1, \varphi=45$. Further numerical details are reported in Appendixes.
  }
  \label{fig:stats}
\end{figure}

\subsection{Quantitative test using numerical simulations in one space dimension.}
To characterize the transition to community formation, 
we simulate $N=100$ species on a domain of length $L_x=30$, fixing the density per
species at $\rho_0=\brho$. We explore the phase diagrams of Figs.~\ref{fig:phasediag_m0},~\ref{fig:phasediag_mfinite} by varying the values of $(m,\sigma^2)$. For each point in the phase diagram, we perform $K=15$ simulations and build the
single-strain density distribution $p(\rho_\mu)$ and the total-density distribution $p(\rho_{\rm t})$, as detailed in App.~\ref{app:1dsimulations}. 
As in Sec.~\ref{sec:2dsimul}, the presence of multiple peaks in $p(\rho_{\rm t})$ and $p(\rho_\mu)$ corresponds to phase separation for the total density and for individual strains, respectively. We thus use the standard deviation of these distributions as an indicator for phase separation. We normalize the standard deviations of the total and single-strain densities such that $\Delta\rho_{\rm t}=1$ and $\langle \Delta\rho_\mu\rangle=1$ correspond to the Poissonian fluctuations of a homogeneous system. 
The shaded regions in the panels of Fig.~\ref{fig:stats} correspond to phase separated simulations, defined as $\langle \Delta\rho_\mu\rangle > 2$. The transitions to community formation observed numerically coincide very well with the dot-dashed instability lines predicted by the theory. 

To distinguish between gregarious and fragmented colonies, we also measured the fragmentation parameter $ F $ in inhomogeneous systems. 
We characterize a system by the most probable value $\bar F$ extracted from the numerical measurement of the distribution $p(F)$ (See Fig.~\ref{fig:histograms}). 
In the absence of average inhibition ($m=0$), we always observe fragmented ecosystems corresponding to $F=1$ (Fig.~\ref{fig:stats}c). In contrast, when $m<0$, 
$\bar F$ increases with $\sigma^2$ as the system transitions from a gregarious colony into a fragmented ecosystem (Fig.~\ref{fig:stats}f). 
The cross-over to $F\simeq 1$ observed numerically corresponds well to the vertical dashed line, which  indicates the BBP transition where our theory predicts that the most unstable mode corresponds to a fully fragmented ecosystem. Our microscopic simulations thus support the theoretical picture derived in
Sec.~\ref{sec:theory} by applying random-matrix theory to the hydrodynamic description of the system.

\subsection{Non-reciprocal interactions.}

\begin{figure*}
  \begin{center}
      \includegraphics[width=\textwidth]{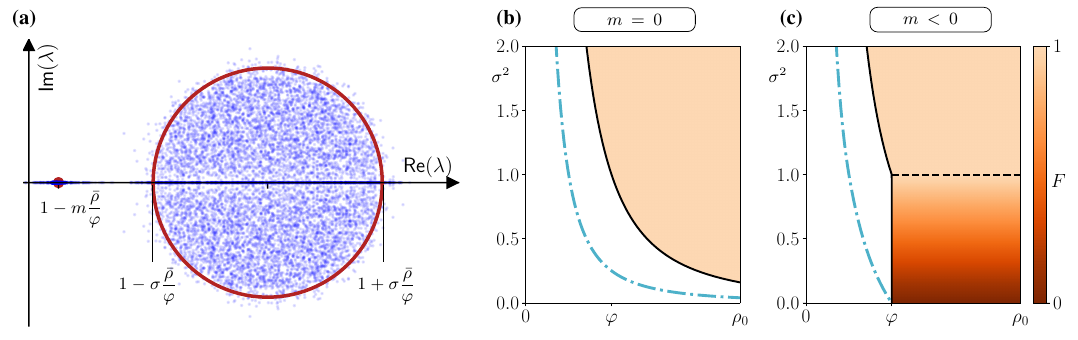}
  \end{center}
    \caption{{Spectral density and phase diagrams in the presence of asymmetric interactions.} ({\bf a}) Spectral distribution of the diffusivity matrix $\cD_{\mu\nu}/\cD_0$ in the complex plane for asymmetric interaction coefficients. We plot both the results from theory (red line and isolated point) and numerical sampling from $K=200$ matrices of size $50\times 50$. For small heterogeneity $(\sigma^2 < m^2)$ and systematic inhibition $m<0$, the spectral density consists the Girko-Ginibre circle plus an outlier. At large heterogeneity ($\sigma^2>m^2$) the outlier is absorbed inside the Girko-Ginibre circle. (\textbf{b}-\textbf{c}) Phase diagrams in the $(\rho_0,\sigma^2)$-space for $\rho_0=\brho$, in the absence ({\bf b}, $m=0$) and in the presence ({\bf c}, $m=-1$) of average motility inhibition. In the shaded region, corresponding to the linear instability, the heatmap represents the fragmentation parameter $F$. To compare with the case of symmetric interactions, we report as a dashed-dotted teal line the spinodal line corresponding to this latter case.}
    \label{fig:asymm}
\end{figure*}

To elucidate the impact of asymmetric interactions between  strains, we consider the case Corr($\kappa_{\mu\nu}$,$\kappa_{\nu\mu}$)$=0$ so that $\kappa_{\mu\nu}$ is generically asymmetric. As in Sec.~\ref{sec:fragmentation}, we work in the regime where $\rho_0=\brho$. We denote by $\lambda_0$ the eigenvalue of $\cD_{\mu\nu}$ with the smallest real part and by $\bfhe_0$ its associated eigenvector. 

In the absence of average motility inhibition ($m=0$), the spectral density of the diffusion matrix given in Eq.~\eqref{eq:Diff_mat} is now described by the Girko-Ginibre ensemble~\cite{ginibre1965statistical}. As shown in Fig.~\ref{fig:asymm}a, it amounts to a uniform disk in the complex plane, centered at $\lambda=(\cD_0, 0)$ and of radius $\cD_0 \rho_0 \sigma/\varphi$. A homogeneous system then becomes unstable when:
\begin{equation}
\text{Re}(\lambda_0)<0 \Leftrightarrow    \sigma^2 > \frac{\varphi^2}{\rho_0^2}  \;.
    \label{eq:inst_fragmented_asymm}
\end{equation}
The corresponding phase diagram is compared in Fig.~\ref{fig:asymm}b with the transition line of the symmetric case. Note that asymmetry stabilizes the homogeneous phase by increasing the instability threshold for $\sigma^2$, consistently with what was reported for other dynamical transitions induced by random interactions~\cite{bascompte2006asymmetric,bunin2017ecological,roy2019numerical,parkavousi2025enhanced}. Finally, in the large-$N$ limit, the fragmentation parameter associated to the smallest eigenvalue can be computed using Eq.~\eqref{eq:F_parameter} and random matrix theory~\cite{benaych2012singular}. It is equal to $F=1$ and confirms the emergence of a fragmented phase.

Next, we turn to the case with average motility inhibition $m<0$. Similarly to the symmetric case, an outlier now appears on the real axis~\cite{tao2013outliers,wurfel2024mean}, located at $\lambda_0 = \cD_0 \left(1 + \frac{\rho_0}{\varphi} m \right)$, see Fig.~\ref{fig:asymm}a. However, at odds with the symmetric case, the location of $\lambda_0$ does not depend on $\sigma^2$. When the instability is induced by the outlier ($\sigma^2<m^2$), the threshold is controlled exclusively by $m$ through:
\begin{equation}
    m < -\frac{\varphi}{\rho_0} \;.
\end{equation}
The corresponding fragmentation parameter is then given by $F=\sigma^2/m^2$~\cite{benaych2012singular}, as in the symmetric case. Therefore, self-inhibition promotes the formation of a gregarious phase for $\sigma^2<m^2$, while at $\sigma^2=m^2$ the BBP transition takes place and induces the fragmentation of the ecosystem, with $F$ becoming equal to $1$. The resulting phase diagram is reported in Fig.~\ref{fig:asymm}c, together with the spinodal line for the symmetric case, reported in teal. All in all, the presence of non-reciprocal couplings in the diffusivity matrix $\cD_{\mu\nu}$ does not change the qualitative phase behavior of the ecosystem, which still displays a transition into a fragmented phase when the heterogeneity $\sigma^2$ becomes sufficiently large.

A notable feature induced by non-reciprocal interactions is the emergence of dynamical phases. Deep in the fragmented phase, we indeed expect a wide range of eigenvalues in the Girko-Ginibre ensemble to become unstable. As these modes have a non-zero imaginary part, the behavior of the system is enriched by the presence of dynamical patterns, as shown in Movie S4-S5. Note that the measurements of the fragmentation parameter in the presence of dynamical patterns still show ecosystem fragmentation, as $F(\bfr)$ is consistently observed to be peaked around $1$ in our simulations. 
Finally, we note that, while symmetric interactions lead to complete phase separation, coarsening may be arrested in the presence of non-reciprocal interactions~\cite{frohoff2021suppression}. 
Investigating how the interaction radius $\ell_{\rm int}$ then plays a role in controlling the size of emergent patterns is an interesting direction for future research.

\section{Extension to other motility-regulation mechanisms}
\label{sec:extension}
Let us now discuss how our results generalize beyond the model defined in Eqs.~\eqref{eq:v_mu_random} and~\eqref{eq:kappa_munu}.

First, we note that our analysis directly extends to other motility-regulation functions beyond Eq.~\eqref{eq:v_mu_random}, such as $v_\mu(\{\rho_\mu\})= v_0
\prod_\nu \phi_{\mu\nu}(\rho_\nu)$. 
Here, $\phi_{\mu\nu} \equiv \exp\{\kappa_{\mu\nu} \cS[(\rho_\nu-\brho)/\varphi]\}$ represents a specific motility-regulation
pathway that couples the strains $\mu$ and $\nu$, leading to an overall factorized regulation of the motility.
As shown in App.~\ref{app:factorized}, random motility regulation again leads to the emergence of a fragmented ecosystem and the change of regulating function mostly leads to different morphologies for the phase diagrams.

A second important ingredient of our model is how heterogeneity is taken into account in the motility regulation mechanism. Instead of the independent metabolic interactions considered above, which leads to i.i.d.~$\kappa_{\mu\nu}$, one can also consider cases where motility is regulated by a single signaling field. We denote by $\gamma_\nu$ the production rate of the corresponding molecule by strain $\nu$ and by 
$\beta_\mu$ the susceptibility of strain $\mu$, so that $\kappa_{\mu\nu} = \beta_\mu \gamma_\nu $ becomes a rank-one matrix. We refer to this case as the single-pathway motility-regulation network.

\begin{figure}
    \begin{center}
        \includegraphics[width=\columnwidth]{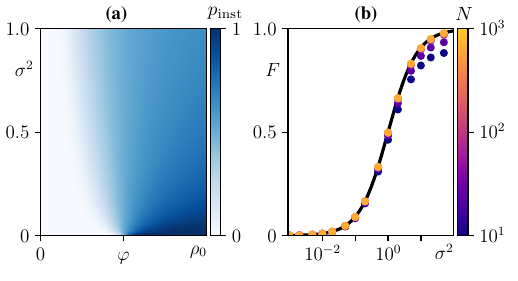}
    \end{center}
    \caption{Single-pathway motility-regulation network for $\brho=\rho_0$. ({\bf a}) Heatmap of the probability $p_{\rm inst}$ for a homogeneous system to be unstable in the $(\rho_0,\sigma^2)$ space. At large heterogeneity $\sigma^2 \equiv \sigma_\beta^2 = \sigma_\gamma^2$, $p_{\rm inst}$ always reaches the asymptotic value $1/2$. ({\bf b}) Fragmentation parameter $F$ obtained from numerical diagonalization of rank-one random matrices (points) versus the theoretical prediction of Eq.~\eqref{eq:F_rankone} (solid line). Data points are colored according to the number of strains $N$ in the system. As predicted by Eq.~\eqref{eq:F_rankone}, $F$ increases with $\sigma^2$: Random interactions again promote community formation. 
    Common parameters: $\bbeta=-1, \bgamma=1$, $\rho_0=\brho$, $\varphi=10$. In (b), $\brho=30$.}
    \label{fig:rank1}
\end{figure}

In Fig.~\ref{fig:rank1}, we consider ecosystems where $\gamma_\nu$ and $\beta_\mu$ are distributed according to: 
\begin{equation}
    \gamma_\nu \sim \cN(\bgamma/N, \sigma_\gamma^2/N),\quad \beta_\mu \sim \cN(\bbeta, \sigma_\beta^2)\;,
\end{equation}
where all coefficients $\bgamma, \bbeta, \sigma^2_{\beta,\gamma}$ are taken to be $\cO(1)$, so that the production rate is small while the strain susceptibility is $~\cO(1)$. When the system is gregarious at $\sigma_\beta^2=\sigma_\gamma^2=0$, which corresponds to $\bgamma \bbeta<-\varphi/\brho$, our numerical simulations and theoretical computations detailed in Appendix~\ref{app:low-rank} show a transition into a fragmented phase upon increasing $\sigma_\beta^2$, characterized by a fragmentation order parameter that can be approximated as
\begin{equation}
F\simeq \frac{\sigma_\beta^2}{\sigma_\beta^2 +\bbeta^2} \;.
\label{eq:F_rankone}
\end{equation}
Increasing the heterogeneity $\sigma^2_\beta$ of motility regulation thus again induces a fragmentation transition.

The topology of the motility-regulation network however induces an important difference with the metabolic case considered above. First, we predict a \textit{finite} probability for a linear instability to occur for any $\sigma_{\gamma,\beta}^2\neq 0$ (see Fig.~\ref{fig:rank1}a and App.~\ref{app:low-rank}). 
Then, the transition from gregarious to fragmented phase is replaced by a smooth crossover from $F=0$ at $\sigma_\beta^2=0$ to $F=1$ as $\sigma_\beta^2 \to \infty$, as shown in Fig.~\ref{fig:rank1}b. Note that this phenomenology extends to the case of $K\ll N$ signaling fields, see App.~\ref{app:low-rank}.

Then, throughout this article, we have considered the case in which the interactions and their heterogeneity scale as $\mathcal{O}(N^{-1})$. Biologically, this is motivated by the requirement to keep interactions finite in the limit of a large number of species. Mathematically, our results show that the case considered here is the most interesting scaling to study ecosystem fragmentation: The transition happens for $\sigma^2\sim \mathcal{O}(1)$ and other scalings can be recovered by taking $\sigma^2 \to \infty$ and $\sigma^2 \to 0$, respectively (See App.~\ref{app:scaling}).

Finally, we have studied a motility-regulation mechanism in which the speed of a strain depends on the local density of its peers through Eq.~\eqref{eq:v_mu_random}. Bacteria can also adapt their motilities to the local gradients of signaling fields, \textit{e.g.} via chemotaxis. This is known to lead to a wealth of pattern-forming mechanisms~\cite{budrene_complex_1991,ben1995complex,brenner1998physical} and it would thus be interesting to
extend our results to such interactions. We expect our formalism to generalize straightforwardly to this case since the coarse-graining of such interactions lead to hydrodynamic equations very similar to Eq.~\eqref{eq:hydro}~\cite{obyrne_lamellar_2020,dinelli2024fluctuating}.

\section{Discussion}
In this work we have shown how random motility regulation promotes
the formation of spatially distinct communities in bacterial ecosystems.
To this end, we have
studied a model comprising $N \gg 1$ strains of run-and-tumble bacteria interacting via motility regulation. Using a combination of microscopic simulations and random matrix theory, we have shown that increasing the heterogeneity in the interaction network leads to the
formation of segregated communities. This occurs both in the
presence and in the absence of an average motility inhibition, and the
phenomenology is robust with respect to the details of motility regulation. All in all, our work thus highlights the importance of motility regulation in the
formation of communities within a microbial ecosystem, even when such a regulation is weak. While characterizing variations in bacterial motility is notoriously difficult, a lot of experimental progress has been made recently~\cite{kurzthaler2024characterization,figueroa20203d} and we hope that our work will thus trigger further exploration of motility regulation in complex bacterial mixtures.

High-dimensional ecosystems are characterized by complex networks of interactions and we have argued that the emergence of fragmented ecosystems could arise either from direct motility regulation due to quorum sensing or from indirect regulation due to metabolic interactions. 
While we expect the latter to be most relevant in nature, the former offers an experimental platform to test our predictions in the future, using recent advances in engineered bacterial systems~\cite{liu_sequential_2011,curatolo_cooperative_2020}. Furthermore, we note that signal-mediated interactions between microbes also play a role in complex environments such as turbulent flows in marine ecosystems~\cite{brumley2020cutting}. This suggests that signalling-based motility regulation could be important across a variety of environmental conditions.

Our results were established using a model that
accounts for motility regulation using a density-based control of the bacteria swimming speed. We note that identical results could be obtained using models in which the tumble rates and durations  depend on the local density, since they lead to hydrodynamic theory similar to Eq.~\eqref{eq:hydro}~\cite{dinelli2024fluctuating}. Furthermore, it is known that bacteria can also adapt their motility based on the gradients of some chemical
fields, \textit{e.g.} via chemotaxis. It would thus be interesting to extend our study to chemotactic
interactions~\cite{budrene_complex_1991,obyrne_lamellar_2020} that are amenable to similar treatments.
In addition, we have here focused on the self-organization that emerges from motility regulation. External regulation of the motility due to inhomogeneous environmental factors can also drive the spatial organization of motile species. How both effects interplay is an interesting question for future work.

In our model, we make the hypothesis of quenched interactions, \textit{i.e.} we 
assume that the inter-strain couplings $\kappa_{\mu\nu}$ do not evolve in
time. However, if weak motility regulation stems from metabolic interactions, it is plausible that the resulting
interaction matrix fluctuates in time. It would thus be interesting to
extend our work to the annealed case, \textit{i.e.} to account for the dynamics of
the signalling molecules themselves~\cite{colombo2023pulsed}, resulting in a time-dependent interaction matrix~\cite{suweis2024generalized,altieri2025unveiling}. Furthermore, we have considered equal strain compositions for the sake of simplicity. Random-matrix theory suggests that fluctuations in the species abundance could impact the phase equilibrium~\cite{thewes2023composition}, which offers an interesting perspective for future work.

Finally, we have focused here on motility-induced community formation and shown how fragmentation may occur on relatively short time-scales due to active transport. On longer time scales, population dynamics will become relevant and other types of interactions, such as
reproduction, cooperation, and competition for resources, will make the accessible phenomenology even richer~\cite{may1972will,lavrentovich2014asymmetric,bunin2017ecological}.

\section*{Acknowledgments} 
The authors thank Victor Chardès, Jeff Gore, Hyunseok Lee, Yizhou Liu, and Peter Sollich for insightful discussions. 
A.A. acknowledges the support of ANR JCJC “SIDECAR” Grant No. ANR-23-CE30-0012-01. J.T. and A.D. acknowledge the support of ANR THEMA through Grant No. ANR-20-CE30-0031 and MIT MISTI GSF grant. A.D. acknowledges funding from an IDEX doctoral fellowship and by the Swiss National Science Foundation through Grant No. 200020E$\_$219164.

\section*{Data availability}
The data that support the findings of this article are publicly available~\cite{data}.

%

\appendix

\section{Motility regulation from metabolic interactions}
\label{app:motility_regulation}
In this Appendix we illustrate how metabolic interactions may lead to Eqs.~\eqref{eq:metabolic_level} and~\eqref{eq:v_mu_random}. Let the metabolic state $\cI_{i, \mu}$ for particle $i$ of species $\mu$ be a function of $n$ chemical fields $\{c_n (\bfr_{i,\mu})\}$. The metabolic state represents the energy available for the bacterium to perform all its biological functions, including swimming. For \textit{E. Coli}, around 5 \% of the available energy is used for motility~\cite{schavemaker2022flagellar}, so that minute fluctuations of $\cI_{i, \mu}$ can have large impact on swimming. This suggests using a sigmoidal function to relate $\cI_{i, \mu}$ to the swimming speed $v_\mu$, as in Eq.~\eqref{eq:v_mu_random}. We note that such a non-linear dependence of motility on the concentration of signalling molecules has been measured for quorum-sensing interactions~\cite{liu_sequential_2011}. Let us now turn to relating $\cI_{i, \mu}$ to the species density fields.

A variety of interactions and signaling mechanisms have been employed to model interactions between motile organisms~\cite{smith2003animal,brumley2020cutting,colombo2023pulsed}. Here we focus on the simplest of cases in which the field $c_n$ is degraded at rate $\lambda_n$, diffuses with diffusivity $D_n$, and is produced by species $\nu$ at rate $\gamma_{n, \nu}$. This leads to the dynamics: 
\begin{equation}
    \partial_t c_n = D_n \nabla^2 c_n - \lambda_n c_n + \sum_\nu \gamma_{n, \nu} \rho_\nu \;.
\end{equation}
The diffusivity of chemicals being very large, we employ a fast-variable approximation, $\partial_t c_n \simeq 0$, leading to:
\begin{equation}
    c_n = K_n \ast \sum_\nu \gamma_{n,\nu} \rho_\nu \;,
\end{equation}
where $K_n$ is the Green's function associated with the screened Poisson equation, and $\ast$ denotes the convolution product.

Within linear response around a homogeneous state $\rho_\nu = \brho$, the metabolic states satisfy:
\begin{eqnarray*}
    \cI_{\mu}(\{c_n(\bfr_{i,\mu}, [\{\rho_\nu\}])\}) &\simeq& \cI_{\mu}(\{c_n(\bfr_{i,\mu}, [\{\brho\}]\}) 
    +\notag\\[0.2cm] 
     &&\sum_{n,\nu} \frac{\partial \cI_{\mu}}{\partial c_n} \gamma_{n,\nu} K_n \ast (\rho_\nu - \brho) \;.
\end{eqnarray*}
Assuming for simplicity $K_n \simeq K$, absorbing their normalization into $\frac{\partial \cI_{\mu}}{\partial c_n}$, and defining:
\begin{equation}
    \kappa_{\mu\nu} = \sum_{n} \frac{\partial \cI_{\mu}}{\partial c_n} \gamma_{n,\nu} \;, \quad \cI_{i,\mu}^0 =  \cI_{\mu}(\{c_n(\bfr_{i,\mu}, [\{\brho\}]\})
\end{equation}
leads to Eq.~\eqref{eq:metabolic_level} of the main text. Note that, in the limit of a single signaling pathway ($n=1$), the interaction matrix $\kappa_{\mu\nu}$ becomes rank-one, so that we recover the single-pathway motility regulation model described in Sec.~\ref{sec:extension} with susceptibility $\beta_\mu = \partial \cI_\mu/\partial c$. 
On the contrary, when $n \sim O(S)$, we recover a random-motility regulation model where $\kappa_{\mu\nu}$ has rank $O(S)$.

\section{Microscopic simulation algorithm.}
\label{app:simulations}
Here we describe the algorithm used to perform particle-based simulations in $1d$ and $2d$ continuous space. In the following, the notation $(i,\mu)$ denotes particle $i$ of strain $\mu$, $\bfr_{i,\mu}$ its position and $\bfu_{i,\mu}$ its orientation.

We initialize the system in a homogeneous configuration at time $t=0$, and, for each particle, we draw the next tumbling time $t_{i,\mu}$ from the exponential distribution $p_{\rm t}(t_i) \equiv \exp(-t_i/\tau)/\tau$, where $\tau$ is the persistence time. Particle positions and orientations between times $t$ and $t + \rmd t$ are then updated as follows. At each time step, we first measure the coarse-grained density fields $\{\tilde\rho_{i,\nu}\}$ as  $\tilde\rho_{i,\nu} \equiv \sum_{(j,\nu)} K(|\bfr_{i,\mu} - \bfr_{j,\nu}|)$, where $K(r)$ is a normalized symmetric kernel given by $K(r) = Z^{-1} \exp\bigl[ -\frac{r^2}{\ell_{\rm int}^2-r^2} \bigr] \Theta(r-\ell_{\rm int})$. Throughout our work, $\ell_{\rm int} = 1$.
We then determine the instantaneous self-propulsion speeds as $v_{i,\mu} = v(\{\tilde\rho_{i,\nu}\})$.

Next, for each particle we check whether the next tumbling event $t_{i,\mu}$ occurs in $[t, t + \rmd t)$. If so, we first evolve the particle position $\bfr_{i\mu}$ in the interval $\rmd t' = t_{i,\mu}-t$ as $\bfr_{i,\mu}(t+\rmd t') = \bfr_{i,\mu}(t) + v(\{\tilde\rho_{i,\nu}\}) \bfu_{i\mu} \rmd t'$. We then draw a new orientation $\bfu_{i\mu}'$ uniformly at random, and update the next tumbling time as $t_{i,\mu} \to t_{i,\mu} + \Delta t_{i,\mu}$, where $\Delta t_{i,\mu}$ is drawn from the exponential distribution $p_{\rm t}$. We then iterate this process until the particle's position has been evolved up to time $t + \rmd t$.

\bigskip

\section{Analysis of $1d$ simulations.}
\label{app:1dsimulations}
To test our theoretical predictions shown in Figs.~\ref{fig:phasediag_m0},~\ref{fig:phasediag_mfinite}, we use simulations of one-dimensional systems, both in the presence of average
motility inhibition ($m = -1$) and in its absence ($m = 0$). We simulate $N=100$ species on a domain of
length $L_x=30$, setting the bare self-propulsion speed to $v_0=1$,
the persistence time $\tau=1$, and the interaction radius $\ell_{\rm int}=1$. We
fix the density per species at $\rho_0=\brho=30$ and choose
$\varphi=20$ for $m = 0$, $\varphi=45$ for $m=-1$. For each value of $\sigma^2$, we perform
$K=15$ simulations. To analyze the resulting data, we divide the space
in boxes of width $\rmd x = 1$ and, for each box $i$, we build a
vector of local strain densities $\bfrho^{(i)} =
(\rho_1^{(i)},\dots,\rho_N^{(i)})$. We then define
the (rescaled) total density $\rho_{\rm t}^{(i)} = N^{-1} \sum_\mu
\rho_\mu^{(i)}$ inside box $i$.
By
averaging over several time samples, we then build the
histogram for each species' density $p(\rho_\mu)$ and the total-density histogram $p(\rho_{\rm
  t})$. 

To estimate the local fragmentation parameter $F^{(i)}$ in each box $i$, we first build a smoothened density profile by averaging $\bfrho^{(i)} \to \langle \bfrho\rangle^{(i)}$ over a time window of $T_{\rm avg} = 10$, sampled every $\rmd t_{\rm sample} = 1$. The local fragmentation is then measured as $F^{(i)} =
1 - N^{-1} \left(\sum_\mu \langle\rho\rangle_\mu^{(i)}\right)^2/|\langle\bfrho\rangle^{(i)}|^2$, from which we build the fragmentation-parameter histogram $p(F)$ by averaging over space and time.

To distinguish between different phases we then compute the following quantities:
\begin{eqnarray}
  \label{eq:deltarho_t}
  \Delta \rho_{\rm t} &\equiv& \sqrt{\frac{\text{Var}[\rho_{\rm t}]}{\rho_0/(N \rmd x)}} \;,
\end{eqnarray}
\begin{eqnarray}
  \label{eq:deltarho_avg}
  \langle \Delta \rho_\mu \rangle &\equiv& \frac{1}{N} \sum_{\mu=1}^N \sqrt{\frac{\text{Var}[\rho_\mu]}{\rho_0/\rmd x}} \;,
\end{eqnarray}
\begin{eqnarray}
  \label{eq:cos2_max}
  \bar F &\equiv& \arg\max [p(F)] \;,
\end{eqnarray}
where the variances Var[$\cdot$] are computed  using the corresponding histogram. 
In the absence of phase separation, in both single-species and total-density histograms the variance is dictated by Poissonian fluctuations, so that both
$\Delta \rho_{\rm t}$ and $\langle \Delta \rho_\mu \rangle$ should be close to $1$. Therefore, deviations of these quantities from $1$ indicate, respectively, the occurrence of phase-separation for the total density and for each species separately. We then use $\Delta \rho_{\rm t} > 2$ as a criterion to distinguish between homogeneous and phase-separated profiles. 

For the simulations reported Fig.~\ref{fig:stats}, we obtain the green disks by averaging $\bar F$ over phase-separated simulations. Using these values we build the heatmap for $\bar F$ in Fig.~\ref{fig:stats}c,e using the same colorcode as in Figs.~\ref{fig:phasediag_m0},~\ref{fig:phasediag_mfinite}. Between two data points, colors are obtained by linear interpolation of our numerical data.

\bigskip

\section{Analysis of $2d$ simulations}
\label{app:heatmps}
In this Appendix we provide details on data analysis for $2d$ simulations corresponding to Figs.~\ref{fig:snapshots},~\ref{fig:histograms}, and complement them with corresponding heatmaps of single-species densities, total density and fragmentation parameters.

Simulations in $2d$ are performed over a periodic domain of size $20 \times 20$, fixing the average density per species to $\rho_0 = 15$ and setting $\bar\rho = 15, \; \varphi = 10, \; v_0 =1, \; \tau = 1$, $\ell_{\rm int} = 1$. 
To build the histograms of Fig.~\ref{fig:histograms}, we sample the density fields and the fragmentation parameter locally over boxes of area $1$ every $\rmd t_{\rm sample} = 4$. At each sampling step we build a histogram for each of these quantities, which we then average over a total time $T_{\rm f} = 2000$ to obtain Fig.~\ref{fig:histograms}.

\begin{center}
    \begin{figure*}[htp!]
        \begin{center}
            \includegraphics[width=\textwidth]{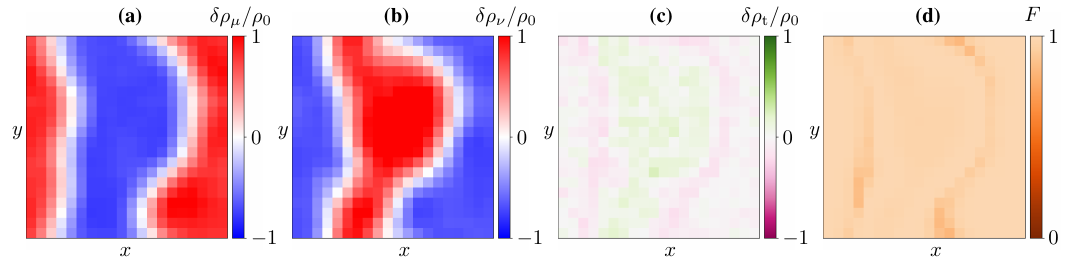}
        \end{center}
        \caption{{Fragmented phase in the absence of average inhibition}, corresponding to panel~$\pentagon$ of Fig.~\ref{fig:snapshots} of the main text. Panels (a, b) show the spatial fluctuations of the density of two representative strains. Panels (c, d) represent the fluctuations of the total-density field (c) and the  fragmentation-parameter field (d), respectively. The total density is homogeneous and the fragmentation parameter $F \simeq 1$, typical of a demixed phase.}
        \label{fig:heatmap1}
        \vspace{0.3cm}
        \begin{center}
            \includegraphics[width=\textwidth]{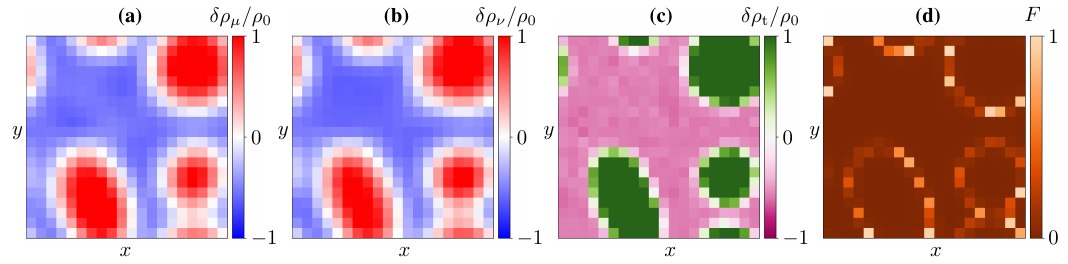}
        \end{center}
        \caption{{Gregarious phase in the presence of average inhibition without heterogeneity}, corresponding to panel~{\scriptsize \faStar[regular]} of Fig.~\ref{fig:snapshots} of the main text. Panels (a, b) show the spatial fluctuations of the density of two representative strains. Panels (c, d) represent the fluctuations of the total-density field (c) and the  fragmentation-parameter field (d), respectively. The total-density field shows a clear phase separation, which is reproduced by each strain. The fragmentation parameter $F \simeq 0$ is typical of a gregarious phase.}
        \label{fig:heatmap2}
        \vspace{0.3cm}
        \begin{center}
            \includegraphics[width=\textwidth]{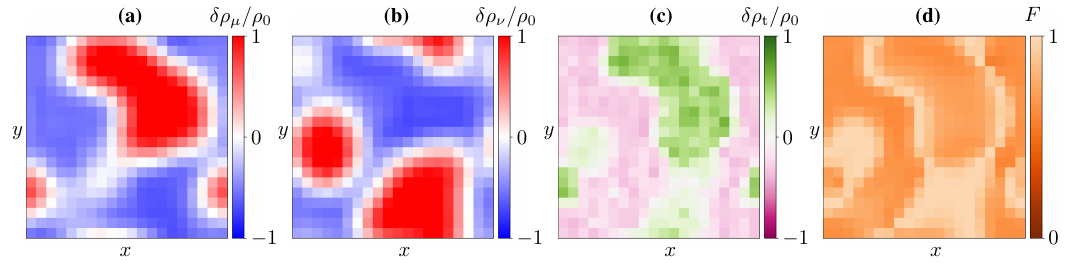}
        \end{center}
        \caption{{Community formation in the presence of both average inhibition and moderate heterogeneity}, corresponding to panel~$\spadesuit$ of Fig.~\ref{fig:snapshots} of the main text. Panels (a, b) show the spatial fluctuations of the density of two representative strains. Panels (c, d) represent the fluctuations of the total-density field (c) and the  fragmentation-parameter field (d), respectively. The density heatmaps show the spatial coexistence between phase-separated ($\delta\rho_t \sim \mathcal{O}(\rho_0)$) and demixed ($\delta\rho_t \simeq 0$, $\delta\rho_\alpha \sim \mathcal{O}(\rho_0)$) regions. The fragmentation parameters shows that the latter are fully fragmented ($F\simeq 1$), while the former are only partially fragmented ($F \lesssim$ 1).}
        \label{fig:heatmap3}
    \end{figure*}
\end{center}

To gain a better insight on the statistics reported in Fig.~\ref{fig:histograms}, we now complement them with density and fragmentation heatmaps in Figs.~\ref{fig:heatmap1}-\ref{fig:heatmap3}. We consider two representative strains $\mu$ and $\nu$, chosen at random among the 50 strains considered in the simulation shown in Fig.~\ref{fig:snapshots} of the main text.  We measure the density fluctuations $\delta\rho_\alpha \equiv \rho_\alpha-\rho_0$ of strain $\alpha$ with respect to a homogeneous profile at density $\rho_0$. Similarly, for the total density field, we define $\delta\rho_{\rm t} = \frac{1}{N}\sum_{\alpha=1}^N \delta\rho_\alpha$. The local fragmentation parameter $F(x,y)$ is defined as in Eq.~\eqref{eq:F_parameter} of the main text. 
To produce the heatmaps shown in Figs.~\ref{fig:heatmap1}-\ref{fig:heatmap3}, we coarse-grain each field over boxes of linear size $1$, and smoothen the result using a convolution with a Gaussian kernel of standard deviation $r=2$, truncated at $2.5r$.

Figure~\ref{fig:heatmap1} corresponds to panel $\pentagon$ of Fig.~\ref{fig:snapshots} in the main text, which shows a fragmented ecosystem in the absence of average motility inhibition $(\sigma^2=2,m=0)$. The two strains are demixed (a, b), with barely any fluctuations in the total density (c). The fragmentation parameter is very close to $1$ everywhere (d).

Figure~\ref{fig:heatmap2} corresponds to panel {\scriptsize \faStar[regular]} of Fig.~\ref{fig:snapshots} of the main text, which shows a gregarious ecosystem resulting from an average inhibition and a lack of heterogeneity $(\sigma^2=0, m=-1)$. Density fluctuations for different strains colocalize (a, b) and reproduce the fluctuations of the total density field (c). The fragmentation parameter is close to $0$ in the whole system (d).

Finally, Fig.~\ref{fig:heatmap3} corresponds to panel $\spadesuit$ of Fig.~\ref{fig:snapshots} of the main text, which shows a fragmentated ecosystem in the presence of motility inhibition $(\sigma^2=2, m=-1)$. Here, a spatial coexistence between demixed and phase-separated regions is observed, which is typical of moderate values of $\sigma^2$. This can be seen on the total density field (c), which contains regions with $\delta\rho_t\simeq 0$ and $\delta\rho_t\sim \mathcal{O}(\rho_0)$, respectively. Demixed regions ($\delta\rho_t\simeq 0$) have a fragmentation parameter equal to $1$, while the phase-separated regions ($\delta\rho_t\sim \mathcal{O}(\rho_0)$) have a large fragmentation parameter that remains smaller than 1 (d). This spatial coexistence explains the double-peaked structure in the distribution of the fragmentation parameter reported in Fig.~\ref{fig:histograms} of the main text.

\section{Derivation of the macroscopic hydrodynamics}
\label{app:coarsegraining}
Here we report the coarse-graining method to derive the hydrodynamics~\eqref{eq:hydro} from the microscopic dynamics~\eqref{eq:v_mu_random}. Our method directly follows~\cite{cates_when_2013,solon_2018_generalized,dinelli2023nonreciprocity,dinelli2024fluctuating}, where the interested reader can find more details.

We start by noticing that the conserved density fields $\{\rho_{\mu}\}$ are diffusive, meaning that they evolve over a temporal scale $L^2$ where $L$ is the system size. Conversely, the microscopic dynamics of the orientational degrees of freedom are characterized by a finite decorrelation time, or persistence time, $\tau$.
As such, in the thermodynamic limit where $L \to \infty$, one can take an adiabatic approximation and assume the density fields to be frozen over timescales $\Delta t \ll L^2$.  
The $N$-body interacting problem of Eq.~\eqref{eq:hydro} thus becomes, at these short time scales, a single-particle problem where $v_\mu(\bfr_{i,\mu},[\{\rho_{\mu}\}])$ is replaced by a position dependent function $v_{\mu}(\bfr_{i,\mu})$ in a frozen activity landscape. This allows us to write a master equation for the probability $\cP_\mu(\bfr,\bfu, t)$ of finding a particle of strain $\mu$ at position $\bfr$ with orientation $\bfu$ at time $t$. This reads:
\begin{equation}
    \partial_t \cP_\mu = -\nabla_\bfr \cdot [ v_\mu(\bfr) \bfu \cP_\mu ] - \frac{\cP_\mu}{\tau} + \frac{1}{\tau \Omega_d} \int_{\mathbb{S}^{d-1}} \rmd^d \bfu \; \cP_\mu \;,
    \label{eq:master_eq}
\end{equation}
where $\mathbb{S}^{d-1}$ denotes the surface of the $d$-dimensional unit sphere, with surface area $\Omega_{d}$. 
From $\cP_{\mu}$ one can define:
\begin{eqnarray}
    p_\mu(\bfr) &=& \int_{\mathbb{S}^{d-1}}   \cP_\mu(\bfr,\bfu) \rmd^d \bfu \, \;, \\
    \bfm_\mu(\bfr) &=& \int_{\mathbb{S}^{d-1}} \bfu \cP_\mu(\bfr,\bfu)  \rmd^d \> \bfu \;, \\
    \bfQ_\mu(\bfr) &=& \int_{\mathbb{S}^{d-1}} \left(\bfu \otimes \bfu - \frac{\bI}{d} \right) \cP_\mu(\bfr,\bfu)  \rmd^d \bfu \> \;,
\end{eqnarray}
corresponding to the marginalized probability distribution for the particle position, the polar-order parameter, and the nematic-order parameter, respectively. The goal of our coarse-graining procedure is to obtain a closed dynamics for $p_\mu$ by integrating out the orientational degree of freedom.

Integrating Eq.~\eqref{eq:master_eq} with respect to the orientation gives:
\begin{equation}
    \partial_t p_\mu = -\nabla_\bfr \cdot [ v_\mu(\bfr) \bfm_\mu ] \;,
\end{equation}
which depends on the polar-order parameter. We thus multiply Eq.~\eqref{eq:master_eq} by $\bfu$ and integrate over the orientation to obtain the dynamics of $\bfm_\mu$:
\begin{equation}
    \partial_t \bfm_\mu = -\nabla_\bfr \cdot \big[ v_\mu(\bfr) \bigl( \bfQ_\mu + \frac{p_\mu}{d} \bI \bigr) \big] - \tau^{-1} \bfm_\mu \;.
\end{equation}
Similarly, one obtains the explicit dynamics of $\bfQ_\mu$ as:
\begin{equation}
    \partial_t \bfQ_\mu = -\nabla_\bfr \cdot [v_\mu(\bfr) \boldsymbol{\chi}_\mu ] - \tau^{-1} \bfQ_\mu \;,
\end{equation}
where $\boldsymbol{\chi}$ is a rank-three tensor.

To close this hierarchy of equations, we note that polar and nematic order parameters relax over a timescale $\tau$, while $p_\mu$ is a conserved, slow mode that varies over macroscopic length and time scales.
We can thus perform an adiabatic approximation for the higher-order harmonics ($\partial_t \bfm_\mu \simeq 0$, $\partial_t \bfQ_\mu \simeq 0$). 
Next, spatial scale separation allows us to perform a gradient truncation to the hierarchy. 
As a result, we obtain $\bfQ_{\mu} \sim \cO(\nabla_\bfr)$, $\bfm_\mu = -\tau \nabla [v_\mu(\bfr) p_\mu] + \cO(\nabla_\bfr^2)$. To leading order in gradients, $p_\mu$ thus evolves via a Fokker-Planck equation:
\begin{equation}
    p_\mu (\bfr) = - \nabla_\bfr \cdot [ \mathbf{V}_\mu p_\mu - \cD_\mu \nabla_\bfr p_\mu]
    \label{eq:FokkPlanck}
\end{equation}
where $\bfV_\mu = - \tau v_\mu \nabla_\bfr v_\mu/d$ and $\cD_\mu = \tau v_\mu^2/d$. 

We can now associate to Eq.~\eqref{eq:FokkPlanck} an It\={o}-Langevin equation that describes the large-scale diffusive dynamics of the particle. Introducing a mesoscopic timescale $\Delta t$, intermediate between the microscopic persistence timescale $\tau$ and the macroscopic diffusive scale $\sim L^2$, 
we can re-instate the density-dependence in the drift and diffusion coefficients through $v_\mu$. The Langevin dynamics of particle $(i,\mu)$ at this mesoscopic scale thus reads:
\begin{eqnarray}
    \dot{\bfr}_{i,\mu} = &&\bfV_\mu (\bfr_{i,\mu}, [\{\rho_\nu\}]) + \nabla_{\bfr_{i,\mu}} \cD_\mu(\bfr_{i,\mu}, [\{\rho_\nu\}]) \notag \\
    &&+ \sqrt{2 \cD_\mu(\bfr_{i,\mu}, [\{\rho_\nu\}])} \boldsymbol{\eta}_{i,\mu}(t)
    \label{eq:Langevin}
\end{eqnarray}
where $ \boldsymbol{\eta}_{i,\mu}(t)$ is a Gaussian white noise with zero mean and delta-correlations. Eq.~\eqref{eq:Langevin} is the starting point to derive the hydrodynamics of the density fields in Eq.~\eqref{eq:hydro}. We define the density field of strain $\mu$ as:
\begin{equation}
    \hat\rho_\mu(\bfr,t) = \sum_{i=1}^{N_\mu} \delta(\bfr - \bfr_{i,\mu}(t)) \;. 
\end{equation}
We then apply It\={o} calculus to obtain its time evolution, following the method by Dean~\cite{dean_langevin_1996,dinelli2024fluctuating}. This yields:
\begin{equation}
    \partial_t \hat\rho_\mu = \nabla_\bfr \cdot [M_\mu \nabla_\bfr \mathfrak{u}_\mu + \sqrt{2 M_\mu \hat{\rho}_\mu} \bfLL_\mu(\bfr,t)] \;.
\end{equation}
Here, $\mathfrak{u}_\mu = \log (\hat{\rho}_\mu v_\mu)$ is the effective chemical potential, $M_\mu = \cD_\mu \hat{\rho}_\mu$ is the collective mobility and $\bfLL_\mu(\bfr,t)$ is a Gaussian white noise field with zero average and delta correlations in time and space. 
Finally, we use a mean-field approximation to discard fluctuations and evaluate all functions $f$ of the fluctuating fields $\hat\rho_\mu$ at their average values $\rho_\mu = \langle \hat\rho_\mu \rangle$: $f(\hat\rho_\mu) \simeq f(\rho_\mu)$, which leads to Eq.~\eqref{eq:hydro}.

\bigskip

\section{Linearized field theory.}
\label{app:linearized_field_theory}
To leading order in a gradient expansion, we approximate the speed of strain $\mu$ as 
$v_{\mu}(\bfr, [\{\rho_\nu\}] \approx v_\mu(\{\rho_\nu(\bfr)\})$.
We then linearize the mean-field hydrodynamic theory for the density
fields Eq.~\eqref{eq:hydro} around a well-mixed homogeneous profile
$\bfrho_0$ to obtain $N$ coupled diffusion equations for the strains: $\partial_t \rho_\mu =
\cD_{\mu\nu} \nabla_{\bfr}^2 \rho_\nu$, where the diffusivity matrix $\cD$
is given by Eq.~\eqref{eq:general_LSA}. For the motility regulation defined by Eq.~\eqref{eq:v_mu_random}, $\cD_{\mu\nu}$
becomes:
\begin{equation*}
  \cD_{\mu\nu} = \cD_\mu^0\left[\delta_{\mu\nu} + \frac{\rho_0}{\varphi} \kappa_{\mu\nu} \cS'\left(\sum_{\nu=1}^N \kappa_{\mu\nu} \frac{\rho_0-\bar{\rho}}{\varphi}\right)  \right]\;,
  \label{eqapp:general_LSA}
\end{equation*}
with $\cS'(x) = 1/\cosh^2(x)$. Further taking $\rho_0 = \brho$ then gives Eq.~\eqref{eq:Diff_mat}.

\bigskip

\section{Phase diagram for an average bacterial density $\rho_0$ away from the threshold $\brho$ of the motility-regulation function} 
\label{app:numerical_phdiag}
To study the behavior of the system for generic density $\rho_0 \neq \brho$, we develop a numerical method to build the phase diagram in the $(\rho_0,\sigma^2)$-space when the threshold is fixed. We first consider the random-motility-regulation model of the main text,~Eqs.~\eqref{eq:v_mu_random},~\eqref{eq:kappa_munu}, with symmetric interactions: $\kappa_{\mu\nu} = \kappa_{\nu\mu}$. The corresponding diffusion matrix $\cD_{\mu\nu}$ for generic $\rho_0 \neq \brho$ is given by Eq.~\eqref{eqapp:general_LSA}
with $\cS'(x) = 1/\cosh^2(x)$. 

\begin{figure*}
    \begin{center}
        \includegraphics[width=\textwidth]{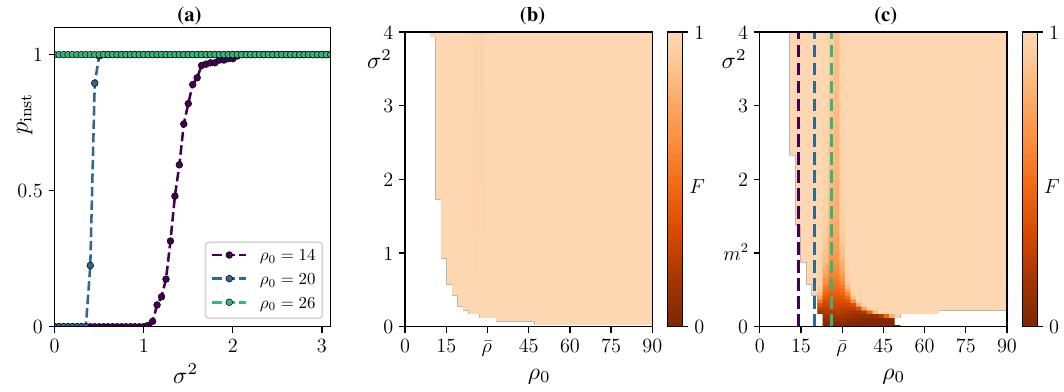}
    \end{center}
    \caption{{Phase diagrams in the $(\rho_0, \sigma^2)$-space for the random-motility-regulation model, defined by Eq.~\eqref{eq:v_mu_random} of the main text.} ({\bf a}): Instability probability $p_{\rm inst}$ as a function of $\sigma^2$ for different values of $\rho_0$ and motility inhibtion $m=-1$. ({\bf b-c}) Phase diagrams in the absence of average motility inhibition (\textbf{b}, $m=0$); in the presence of average motility inhibition (\textbf{c}, $m<0$). White regions represent stable homogeneous phases, while shaded regions correspond to phase-separated states. The color encodes the fragmentation parameter $F$, ranging from $0$ (gregarious ecosystem) to $1$ (fragmented ecosystem). ({\bf c}) Dashed lines correspond to the values of $\rho_0$ of panel a. Parameters: $\varphi=20, \brho=30, v_0=1$.} 
    \label{fig:rho0_phase_diag}
    \vspace{1cm}
    \begin{center}
        \includegraphics[width=\textwidth]{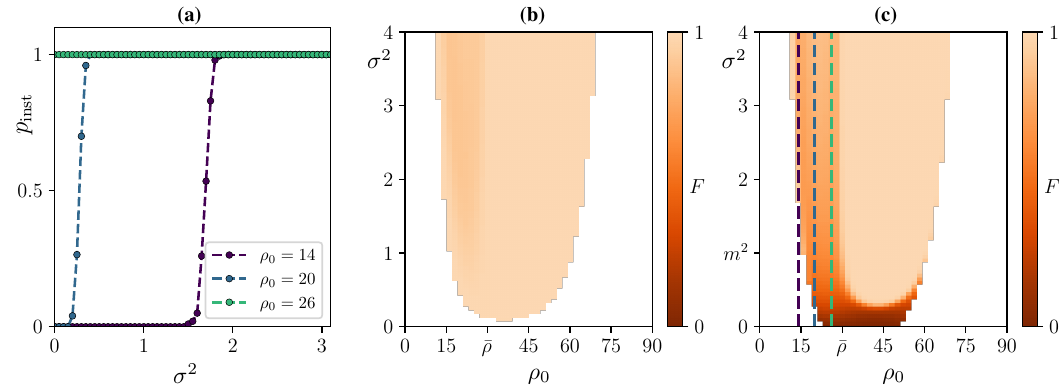}
    \end{center}  
    \caption{{Phase diagrams in the $(\rho_0, \sigma^2)$-space for the factorized-motility-regulation model, defined by~\eqref{SMeq:factorized}.} ({\bf a}): Instability probability $p_{\rm inst}$ as a function of $\sigma^2$ for different values of $\rho_0$ and motility inhibtion $m=-1$. ({\bf b-c}) Phase diagrams in the absence of average motility inhibition (b, $m=0$); in the presence of average motility inhibition (c, $m<0$). White regions represent stable homogeneous phases, while shaded regions correspond to phase-separated states. while shaded regions correspond to phase-separated states. The color encodes the fragmentation parameter $F$, ranging from $0$ (gregarious ecosystem) to $1$ (fragmented ecosystem). ({\bf c}) Dashed lines correspond to the values of $\rho_0$ of panel a. Parameters: $\varphi=20, \brho=30, v_0=1$.} 
    \label{fig:factorized}
\end{figure*}

We use a grid of values $(\rho_0,\sigma^2)$ and, for each point, we determine the probability of a linear instability to occur, $p_{\rm inst}(\rho_0,\sigma^2)$. In the unstable region, we also compute the associated fragmentation parameter $F(\rho_0,\sigma^2)$. To do so, we proceed as follows:
\begin{enumerate}
  \item We generate $K=200$ random matrices $\{\cD^{(k)}_{\mu\nu}\}$ of size $N \times N$ of the form~\eqref{eqapp:general_LSA}, with $N=200$.
  \item We diagonalize each matrix
   $\cD^{(k)}_{\mu\nu}$
    numerically. We determine the eigenvalue $\lambda_0^{(k)}$ with smallest real part and the associated eigenvector $\bfe_0^{(k)}$. 
    If $\text{Re}[\lambda_0^{(k)}]<0$, corresponding to a linear instability, we:
  \begin{itemize}
      \item add the contribution of sample $k$ to the instability probability: $p_{\rm inst} \to p_{\rm inst} + 1/K$.
      \item measure the fragmentation parameter associated with $\bfe_0^{(k)}$:
      \begin{equation}
        F^{(k)} = 1- \left| \bfhe_{0}^{(k)} \cdot \bone \right|^2 = 1- \frac{1}{N} \left| \sum_{\mu=1}^N \hat{e}_{0,\mu}^{(k)} \right|^2 \;.
    \label{eq:herm_cos}
    \end{equation}
    where $|\cdot|$ denotes the complex modulus.
  \end{itemize}
  \item We iterate the previous step for the $K$ samples to obtain the final value for the instability probability $p_{\rm inst}(\rho_0,\sigma^2)$. We then classify $(\rho_0,\sigma^2)$ as a linearly unstable point if:
  \begin{equation}
      p_{\rm inst}(\rho_0,\sigma^2) > \frac{1}{2} \;.
  \end{equation}
  Note that the transition from $p_{\rm inst}(\rho_0,\sigma^2)\simeq 0$ to $p_{\rm inst}(\rho_0,\sigma^2)\simeq 1$ is very sharp so that the position of the transition line is largely insensitive to the precise threshold for $p_{\rm inst}(\rho_0,\sigma^2)$, See Fig.~\ref{fig:rho0_phase_diag}(a).
  \item In the unstable region, the final value for the fragmentation parameter $F(\rho_0,\sigma^2)$ is obtained by averaging $\{F^{(k)}\}$ over all samples where $\lambda_0^{(k)} < 0$.
\end{enumerate}

This procedure allows us to build the phase diagrams shown in Fig.~\ref{fig:rho0_phase_diag}, both in the absence (b) and in the presence (c) of systematic inhibition. In both cases, increasing $\sigma^2$ leads to phase-separation and fragmentation of the ecosystem. Note that heterogeneity-induced fragmentation persists at average densities $\rho_0$ much larger than $\brho$. It naturally disappears in the small density limit, where interactions become negligible. 
All in all, the scenario reported at $\rho_0 = \brho$ in the main text is representative of the transition observed elsewhere in the $(\rho_0,\sigma^2)$ phase diagram.

\section{Factorized motility regulation} 
\label{app:factorized}
We now consider a different motility-regulation mechanism that makes the speed $v_\mu$ of  strain $\mu$ be the product of independent modulations resulting from specific interactions with the other strains:
\begin{equation}
    v_{\mu} = v_0 \prod_{\nu=1}^N \phi_{\mu\nu}[\rho_\nu] \;, \qquad \phi_{\mu\nu} \equiv \exp \left[ \kappa_{\mu\nu} \cS\left( \frac{\tilde\rho_\nu-\brho}{\varphi} \right)\right] \;.
    \label{SMeq:factorized}
\end{equation}
Accordingly, the diffusivity matrix reads:
\begin{equation*}
  \cD_{\mu\nu} = \cD_\mu^0\left[\delta_{\mu\nu} + \frac{\rho_0}{\varphi} \kappa_{\mu\nu} \cS'\left(\frac{\rho_0-\bar{\rho}}{\varphi}\right)  \right]\;.
\end{equation*}
Note that Eq.~\eqref{eq:Diff_mat} follows again from $\rho_0 = \brho$.

We now employ the numerical method of App.~\ref{app:numerical_phdiag} to build the phase diagram of the system in the $(\rho_0,\sigma^2)$-space. The emerging picture is shown in Fig.~\ref{fig:factorized} both in the absence (panel a) and in the presence (b) of systematic motility inhibition $m$. Again, random interactions induce fragmentation. Note that, in this case, the fragmentation transition is sent to larger value of $\sigma^2$ as $\rho_0$ increases, since each regulation pathway has a weaker and weaker impact on the motility as $\rho_0$ increases. 

\section{Low-rank motility-regulation networks}
\label{app:low-rank}
In this Section we first derive the results reported in Fig.~\ref{fig:rank1} of the main text, by showing the emergence of community formation for the single-pathway motility-regulation model at $\rho_0=\brho$. To do so, we compute the instability probability $p_{\rm inst}$ for this model and the associated fragmentation parameter $F$. We discuss all possible scaling regimes, showing that the case discussed in the main text is the marginal case from which all others can be recovered by taking the appropriate limits as $N\to \infty$.

Finally, we extend our results to a $K$-field motility-regulation models, in which bacteria interact via $K$ signaling pathways.

\subsection{Single-pathway motility-regulation model}
We consider a generic single-field regulatory network where each strain $\nu$ produces a chemical $c$ with a rate:
\begin{equation}
  \gamma_\nu = \cN\left(\frac{\bgamma}{N}, \frac{\sigma_\gamma^2}{N}\right) \;.
\end{equation}
The chemical field $c$ then weakly impacts the self-propulsion speed
of strain $\mu$. The response of strain $\mu$ to $c$ is captured by a
susceptibility coefficient:
\begin{equation}
  \beta_\mu = \cN\left(\bbeta, \sigma_\beta^2\right) \;,
  \label{eq:beta-stat}
\end{equation}
so that the interaction matrix is a rank-one matrix given by $\kappa_{\mu\nu} =
\beta_\mu \gamma_\nu$. We then respectively denote the average and variance of $\kappa_{\mu\nu}$  by:
\begin{eqnarray}
    \langle \kappa_{\mu\nu} \rangle &\equiv& \frac{m}{N} = \frac{\bgamma\bbeta}{N} \;, \\ \text{Var}[\kappa_{\mu\nu}] &\equiv& \frac{\nu^2}{N} = \frac{\sigma_\gamma^2 \sigma_\beta^2}{N} \left(1 + \frac{\bbeta^2}{\sigma_\beta^2} + \frac{1}{N}\frac{\bgamma^2}{\sigma_\gamma^2}\right) \;.
\end{eqnarray}
Below we discuss the various scalings of $\bgamma$, $\bbeta$, $\sigma^2_{\gamma,\beta}$ with $N$, assuming that $\frac m N, \frac{\nu^2}N$ can be, at most, $\mathcal{O}(1)$ as $N\to\infty$. 
We note that the results presented in the main text are recovered by taking $\bgamma, \bbeta, \sigma^2_{\gamma,\beta} \sim \cO(1)$.

The linear stability of a homogeneous profile is controlled by the diffusivity matrix $\cD_{\mu\nu}$, which, when $\rho_0=\bar \rho$,  reads $\cD_{\mu\nu} = \cD_0 (\delta_{\mu\nu} + \frac{\rho_0}{\varphi} \kappa_{\mu\nu})$. Without loss of generality, we set $\cD_0 = 1$ by making the appropriate choice of time and space units. We then look for the smallest eigenvalue and the associated eigenvector. The fact that $\kappa_{\mu\nu}$ is rank-1 constrains the spectrum of $\cD_{\mu\nu}$. Direct algebra shows the latter to comprise one positive eigenvalue $\lambda_1=1$ with multiplicity $N-1$ and one outlier eigenvalue $\lambda_0$ with multiplicity 1, where:
\begin{equation}
  \lambda_0 = 1 + \frac{\rho_0}{\varphi} \bfbeta \cdot \bfgamma=  1 +\frac{\rho_0}{\varphi} \sum_{\mu=1}^N \beta_\mu \gamma_\mu \;.
  \label{eq:distr-k0}
\end{equation}
The eigenvector $\bfhe_0$ associated with $\lambda_0$ is the susceptibility vector $\bfbeta = (\beta_1, \dots, \beta_N)$. The only way a linear instability can occur is when $\lambda_0$ becomes negative. 

We can now determine the statistics of $\lambda_0$ and $\bfhe_0$ from their expressions above. In the limit $N\gg 1$,  the central limit theorem tells us that $\lambda_0$ is distributed as:
\begin{equation}
  \lambda_0 =  1 + \frac{\rho_0}{\varphi} \cN\left(m, \nu^2\right) \;.
\end{equation}

We denote by $p_{\rm inst}$ the probability of observing a linear instability for a given choice of parameters, so that $p_{\rm inst} = \text{Prob}[\lambda_0 < 0]$. For a finite value of $N$, $p_{\rm inst}$ is readily computed as:
\begin{equation}
  p_{\rm inst}(\brho,\sigma^2;N) = \int_{-\infty}^{0} \rmd k\frac{1}{\sqrt{2 \pi \nu^2}} \exp\left( -\frac{(k-\frac{\varphi}{\rho_0}-m)^2}{2 \nu^2} \right) \;.
  \label{eq:pinst}
\end{equation}
The behavior of $p_{\rm inst}$ in the large-$N$ limit then depends on the scaling of $m$ and $\nu^2$ when $N\to\infty$, as we now detail:
\begin{enumerate}
    \item When $|m| \to \infty$, the term $\varphi/\rho_0$ becomes negligible in~\eqref{eq:pinst}. Depending on the relative scaling of $m^2$ and $\nu^2$, different regimes are observed:
    \begin{enumerate}
        \item[a.] {When $\nu^2/m^2 \to \infty$: 50\% instability probability.} \\
        $\lambda_0$ becomes Gaussian-distributed with a standard deviation that diverges faster than its mean. Therefore, the probability for $\lambda_0$ to be negative approaches $p_{\rm inst}= 1/2$ as $N \to \infty$.
        \item[b.] {When $\nu^2/m^2 \to 0$: Deterministic limit.} \\
        The distribution of $\lambda_0$ becomes a Dirac delta centered around $\langle \lambda_0 \rangle \sim m \rho_0/\varphi$. Depending on whether $m$ is positive or negative, $p_{\rm inst}$ is thus $0$ or $1$, respectively.
        \item[c.] {When $\nu^2/m^2 \sim \cO(1)$: Finite instability probability, independent of $\rho_0, \varphi$}.\\
        $p_{\rm inst}$ can assume any finite value in $(0,1)$. For $m<0$, the average interactions favor phase separation and $p_{\rm inst}>1/2$, while for $m>0$ the converse occurs and $p_{\rm inst}<1/2$. In this limit, however, the activation threshold $\rho_0/\varphi$ plays no role in determining the conditions for instability, which is purely determined by $m$ and $\nu^2$.
    \end{enumerate}
    \item $m \to 0$. The average of the eigenvalue $\langle \lambda_0 \rangle \to 1$. In this case, the instability is entirely controlled by the fluctuations:
    \begin{enumerate}
        \item[a.] {When $\nu^2 \to \infty$: 50\% instability probability.} \\
        In this limit, $\lambda_0$ becomes Gaussian-distributed around $1$ with a diverging variance, so that $p_{\rm inst} \to  1/2$. 
        \item[b.] {When $\nu^2 \to 0$: The ecosystem is homogeneous}. \\
        $\lambda_0$ is delta-distributed around $1$, leading to $p_{\rm inst} = 0$.
        \item[c.] {When $\nu^2 \sim \cO(1)$: Small, finite instability probability.} \\
        $\lambda_0$ has a Gaussian distribution centered around $1$ with finite variance, so that $p_{\rm inst}$ can assume any finite value in $(0,1/2)$.
    \end{enumerate} \item $m \sim \cO(1)$. The average value of $\lambda_0$ is $\langle \lambda_0 \rangle = 1+m \rho_0/\varphi$. A monodisperse system ($\nu^2=0$) then undergoes an instability with probability $1$ when $\bbeta\bgamma<-\varphi/\rho_0$. Fluctuations then affect the instability condition as follows:
    \begin{enumerate}
        \item[a.] {When $\nu^2 \to \infty$: 50\% instability probability.} \\
        $\lambda_0$ becomes Gaussian-distributed around a finite value $1+m\rho_0/\varphi$ with a diverging variance, hence a linear instability can always occur with probability $p_{\rm inst} \to 1/2$. 
        \item[b.] {When $\nu^2 \to 0$: Deterministic limit.} \\
        The distribution of $\lambda_0$ becomes a Dirac delta $\delta(m-\langle\lambda_0\rangle)$. The instability of the system is exclusively controlled by whether $m$ is larger or smaller than $-\varphi/\rho_0$, so that $p_{\rm inst}$ is either $1$ or $0$.
        \item[c.] {When $\nu^2 \sim \cO(1)$:  Finite instability probability.} \\
        This is the only case in which $\lambda_0$ is Gaussian distributed with both finite mean and variance, corresponding to a finite probability of instability $p_{\rm inst} \in [0,1]$. This is also the case discussed in the main text, where the interplay between all terms $m, \nu^2$ and $\varphi/\rho_0$ determines the instability conditions.
    \end{enumerate}
\end{enumerate}
In conclusion, the choice of the scaling {3.c} corresponds to the marginal case where all phases can be observed with a finite probability. The discussion above shows how, starting from scaling {3.c}, all regimes can indeed be recovered by sending $m, \nu^2$ to $0$ or $\infty$. Scaling 3.c is thus the most relevant case, which legitimates its study in the main text, and we now detail the computation of the fragmentation parameter in this regime.

\subsection{Estimation of the fragmentation parameter in the single-field model}
To study the emergence of random community formation in this model, we
proceed as in the main text and compute the global fragmentation parameter $F$ as a function of $(\sigma^2_\gamma,\sigma^2_\beta, \bbeta, \bgamma)$. Here we focus on the scaling regime \textbf{3c} where $m, \nu^2 \sim \cO(1)$ as $N \to \infty$.

Since the outlier eigenvector satisfies $\bfhe_0 \propto \bfbeta$, $F$ is given by:
\begin{equation}
  F = 1 - \llangle \frac{(\bfbeta \cdot \bone)^2}{|\bfbeta|^2} \rrangle_{\rm inst} \;, 
  \label{eq:F-prodreact}
\end{equation}
where $\langle \cdot \rangle_{\rm inst}$ is the average over realizations of $\bfbeta$
and $\bfgamma$ for which the eigenvalue $\lambda_0$ is negative. In the general case of finite $p_{\rm inst}$, this constraint prevents us from deriving a closed expression for $F$. To proceed, we thus approximate the average in~\eqref{eq:F-prodreact} by an unconstrained average, which we expect to work well for small $\sigma^2_{\gamma,\beta}$ if the system is unstable at $\sigma^2_{\gamma,\beta}=0$. Furthemore, we estimate $F$ by computing the average of
the numerator and the denominator in~\eqref{eq:F-prodreact}
separately and approximate the average of the ratio by the ratio of the averages. Fig.~\ref{fig:rank1}b of the main text shows that both these approximations give a very good estimate of
$F$, when compared against numerics.

We thus compute $F$ as:
\begin{eqnarray}
  F = 1 - \llangle \frac{ \left(\sum_{\mu=1}^N \beta_\mu\right)^2 }{ N  \sum_{\mu=1}^N \beta_\mu^2  } \rrangle \approx 1 - \frac{\llangle \left(\sum_{\mu=1}^N \beta_\mu\right)^2 \rrangle}{ N \llangle \sum_{\mu=1}^N \beta_\mu^2  \rrangle}\;.
  \label{eq:F1-single}
\end{eqnarray}
We then compute separately:
\begin{eqnarray}
  &&\llangle \left(\sum_{\mu=1}^N \beta_\mu\right)^2 \rrangle =  \sum_{\mu=1}^N \llangle \beta_\mu^2 +  \sum_{\nu\neq\mu}  \beta_\mu\beta_\nu \rrangle \notag \\[0.2cm]
  \label{eq:F2-single}
  &&= N \llangle \beta_\mu^2 \rrangle + N (N-1) \langle \beta_\mu \rangle^2 \\[0.2cm]
  &&= N  \left[ {\bbeta}^2 + \sigma_\beta^2\right] + N(N-1) \bbeta^2 = N^2 {\bbeta}^2 + N \sigma_\beta^2  \notag \;,
\end{eqnarray}
and:
\begin{eqnarray}
  N \llangle \sum_{\mu=1}^N \beta_\mu^2  \rrangle =  N^2 \left[ {\bbeta}^2 + \sigma_\beta^2\right] = N^2 {\bbeta^2} + N^2 \sigma_\beta^2\;.
  \label{eq:F3-single}
\end{eqnarray}
Injecting~\eqref{eq:F2-single} and~\eqref{eq:F3-single} into~\eqref{eq:F1-single} then leads to:
\begin{eqnarray}
  F \approx 1 - \dfrac{ \bbeta^2 + {\sigma_\beta^2}/{N}}{\bbeta^2 + {\sigma_\beta^2}} = \dfrac{1-1/N}{1+\bbeta^2/\sigma_\beta^2}\;.
  \label{eq:F-res-prodreact2}
\end{eqnarray}
Equation~(\ref{eq:F-res-prodreact2}) shows that the fragmentation parameter $F$  is entirely controlled by the relative scaling of $\bbeta^2$ and $\sigma_{\beta}^2$ in the $N \to \infty$ regime. This ratio can be tuned freely when $m, \nu^2 \sim \cO(1)$, leading to three distinct regimes as $N \to \infty$:
\begin{itemize}
    \item {When ${\sigma_\beta^2}/{\bbeta^2} \to 0$, the ecosystem is gregarious.}\\
    The susceptibility vector $\bfbeta$ is dominated by its average, which is proportional to $\bone$: the fragmentation parameter is always $0$ and heterogeneity does not impact the gregarious phase.
    \item {When ${\sigma_\beta^2}/{\bbeta^2} \to \infty$, the system undergoes a discontinuous phase transition between gregarious and fragmented states.}\\
    The susceptibility is dominated by its Gaussian fluctuations and the fragmentation parameter  discontinuously jumps from $0$ to $1$ at $\sigma_\beta^2=0$: very small fluctuations are then sufficient to trigger fragmentation. 
    \item {When ${\sigma_\beta^2}/{\bbeta^2} \sim \cO(1)$, the system undergoes a smooth crossover between gregarious and fragmented states.} \\
    The fragmentation parameter is always finite, and smoothly varies between $F=0$ at $\sigma_\beta^2=0$ and $F=1$ at $\sigma_\beta^2 \to \infty$. This is the case studied in the main text, from which all other cases can be retrieved by taking the corresponding limits. 
\end{itemize}

\subsection{$K$-field motility-regulation model} 
We finally extend our previous results to the case of a finite number $K$ of signaling pathways. In this model, each bacterial strain $\nu$ produces $K$ chemicals, labeled by $i \in \{1, \dots, K \}$, with a production rate:
\begin{equation}
    \gamma^{(i)}_\nu \sim \cN(\bar\gamma^{(i)}/N, \sigma_\gamma^2/N) \;.
\end{equation}
As in the single-field case, each strain $\mu$ then adapts their speed in response to chemical $i$ with a susceptibility:
\begin{equation}
    \beta^{(i)}_\mu \sim \cN(\bbeta^{(i)}, \sigma_\beta^2) \;,
\end{equation}
and we define the overall interaction matrix as:
\begin{equation}
    \kappa_{\mu\nu} = \frac{1}{K} \sum_{i=1}^K \kappa_{\mu\nu}^{(i)}\;, \quad \text{where} \quad \kappa_{\mu\nu}^{(i)} = \beta_\mu^{(i)} \gamma_\nu^{(i)} \;.
    \label{eq:K_field-mat}
\end{equation}
For the sake of simplicity, we consider the case where $\bbeta^{(i)}=\bbeta$ and $\bgamma^{(i)}=\bgamma$, and assume $\sigma_{\gamma}^2 = \sigma_\beta^2 = \sigma^2$. We then remark that $\langle \kappa_{\mu\nu} \rangle = m/{N}$ and Var$[\kappa_{\mu\nu}] = {\nu^2}/{N}$ as in the single-field model. As long as all vectors $\{\bfbeta^{(i)}\}$ and all $\{\bfgamma^{(i)}\}$ are, respectively, linearly independent, the interaction matrix $\kappa_{\mu\nu}$ has rank $K$, so that only $K$ eigenvalues can be associated with an unstable mode for the density dynamics. The linear independence of these Gaussian vectors is satisfied almost surely (except on a measure-zero set).

\begin{figure}[htp!]
    \begin{center}
        \includegraphics[width=\columnwidth]{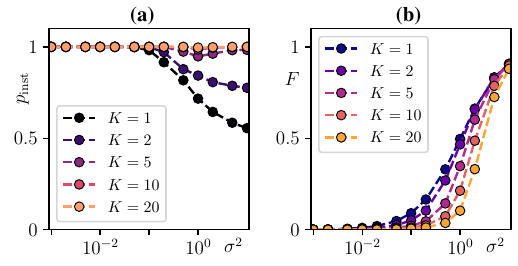}
    \end{center}
    \caption{{$K$-field model: instability probability and fragmentation parameter.} ({\bf a}) Instability probability $p_{\rm inst}$ as a function of $\sigma^2$ and the number of motility-regulation pathways $K$. ({\bf b}) Fragmentation parameter $F(\sigma^2,K)$ for $N=200$ strains. For all panels, each curve is obtained by sampling $M=1000$ matrices of the form~\eqref{eq:K_field-mat}. Common parameters for both panels: $\bbeta=-1, \bgamma=1, \brho=\rho_0=30, \varphi=10$.} 
    \label{SMfig:Kfield}
\end{figure}

Assuming $\rho_0 = \brho$, we determine numerically the condition for which the diffusivity matrix $\cD_{\mu\nu} = \cD_0 (1+\frac{\rho_0}{\varphi} \kappa_{\mu\nu})$ admits a negative eigenvalue. To do so, we use the same protocol as in App.~\ref{app:numerical_phdiag}. This allows us to determine the probability of a linear instability $p_{\rm inst}$ and the fragmentation parameter $F$ for different values of the number of pathways $K$, the number of strains $N$, and the heterogeneity $\sigma^2$. In Fig.~\ref{SMfig:Kfield} we plot our numerical results for an ecosystem of $N=200$ strains in the regime where $m<-\varphi/\rho_0$, corresponding to a linearly unstable monodisperse colony. 
We checked that $N$ is large enough that our results are converged. Fig.~\ref{SMfig:Kfield}a shows the behavior of the instability probability upon increasing $\sigma^2$ and $K$. The instability probability $p_{\rm inst}$ increases systematically with the number of motility-regulation pathways $K$. This suggests that a higher number of motility-regulation pathways typically enhances the tendency of the system to become unstable. Finally, Fig.~\ref{SMfig:Kfield}b reports the behavior of $F$ as $\sigma^2$ is varied for several values of $K$. In all cases, our numerics suggest a smooth crossover from $F=0$ to $F=1$, in agreement with our theory for $K=1$. Upon increasing $K$, the gregarious phase becomes more resistent at large $\sigma^2$.

All in all, the $K$ signaling pathway model confirms that heterogeneity promotes the fragmentation of bacterial colonies. By comparing the phenomenology with that of the `metabolic' random motility-regulation model, we see that there are qualitative differences with how the ecosystem becomes fragmented. For instance, the $K$-signaling pathway model allows for a finite instability probability while the random-motility regulation model always predict $p_{\rm inst}=0$ or $p_{\rm inst}=1$. In that sense, metabolic interactions appear as a more robust community-formation mechanism than targeted motility-regulation, which is a quite surprising result.

\section{Scaling analysis for the random-motility-regulation model}
\label{app:scaling}
In this section, we illustrate explicitly how different scalings of the interaction-matrix entries $\kappa_{\mu\nu}$ affect the collective behavior of the system. In particular, we show that the scaling studied in the main text corresponds to the marginal case, from which all other regimes can be obtained by taking  relevant limits.

We consider a symmetric interaction matrix with Gaussian entries $\kappa_{\mu\nu} = \kappa_{\nu\mu} = \cN(m/N, \sigma^2/N)$, where  $m, \sigma^2$ are not constrained to be $\cO(1)$. For simplicity, we take $\brho=\rho_0$. We denote by $\lambda_0$ the smallest eigenvalue of the diffusivity matrix $\cD_{\mu\nu} = \cD_0 \left( \frac{\varphi}{\rho_0}\delta_{\mu\nu} + \kappa_{\mu\nu} \right)$, and by $\bfhe_0$ the associated eigenvector. To lighten the notation, we use length and time units such that $\cD_0 = 1$. We now discuss the instability condition $\lambda_0 < 0$ for different scaling regimes, both in the absence and in the presence of an average motility inhibition $m$, and comment on the value of the fragmentation parameter:
\begin{equation}
    F=1- \langle[ \bfhe_0 \cdot \bone ]^2\rangle \,
    \;.
\end{equation}
where $\bone = N^{-1/2} (1,\dots,1)$.

\bigskip 

\emph{No average interaction}. In the case where $m=0$, the instability condition in the $N\to \infty$ limit is entirely determined by $\sigma^2$ and $\varphi/\rho_0$.
\begin{enumerate}
    \item {When $\sigma^2 \to 0$: The ecosystem is homogeneous}. \\
    In this limit, the Wigner semicircle degenerates into a $\delta$-distribution centered around $\varphi/\rho_0 > 0$. All eigenvalues of $\cD_{\mu\nu}$ are positive and no linear instability is observed.
    \item {When $\sigma^2 \to \infty$: The ecosystem is fragmented}. \\
    In this limit, the left edge of the Wigner semicircle diverges to $-\infty$ and the system is always linearly unstable with $F=1$.
    \item {When $\sigma^2 \sim \cO(1)$: The system undergoes a phase transition between homogeneous and fragmented states.}\\
    This is the marginal case studied in the main text, where we observe the phase transition between a homogeneous and a fragmented phase for a finite value of $\sigma^2$.
\end{enumerate}
We note that these three regimes are recovered by taking the corresponding limits in Eq.~\eqref{eq:transition1} of the main text.

\bigskip 

\emph{Average inhibition} We now consider the case where $m<0$ and study all possible regimes as $N \to \infty$.
\begin{enumerate}
    \item $m \to \infty$: In this limit, the diagonal contribution $\varphi/\rho_0$ plays a vanishing role as $N$ increases, so that the phase behavior of the system is dictated by $\sigma^2$ and $m$ only. The homogeneous phase is thus always unstable to phase-separation, leading to either a gregarious or a fragmented ecosystem. We thus distinguish between the following sub-cases:
    \begin{enumerate}
        \item[a.] {When $\sigma^2/m^2 \to 0$: The system is gregarious.}\\
        In this case, an outlier eigenvalue is present at $\lambda_0 \sim \frac{\varphi}{\rho_0}m$. It is  always located to the left of the  left edge of the Wigner semicircle, which is at $-2(1+ \frac{\rho_0}{\varphi}\sigma)$, when $\sigma/m \to 0$. Consequently, the most unstable mode is always associated with the outlier, and the corresponding eigenvector has a fragmentation parameter $F = \sigma^2/m^2 \to 0$: the system is  always in the gregarious phase.
        \item[b.] {When $\sigma^2/m^2 \to \infty$: The ecosystem is fragmented}. \\
        In this case, the spectrum is composed of the Wigner semicircle only, whose left edge is located at $\lambda_0 \sim -2 \frac{\rho_0}{\varphi}\sigma$. The eigenmode associated with $\lambda_0$ is thus characterized by $F=1$. As $\sigma$ is diverging, the system is always in the fragmented phase.
        \item[c.] {When $\sigma^2/m^2 \sim \cO(1)$:  The system undergoes a phase transition between homogeneous and fragmented states.} \\
        In this case, the ecosystem undergoes a transition from a gregarious state for $m^2>\sigma^2$, where the outlier $\lambda_0 = \frac{\rho_0}{\varphi} m(1+\sigma^2/m^2)$ 
        is the smallest eigenvalue, to a fragmented state, when $m^2<\sigma^2$. In the gregarious phase, $F=\sigma^2/m^2$, while in the fragmented phase it saturates at $F=1$. We note that, in contrast with the case discussed in the main text, the homogeneous phase is here always unstable. 
    \end{enumerate}
     \item $m \sim \cO(1)$: In this limit, depending on the scaling of $\sigma^2$, all three phases may be observed: homogeneous, gregarious, fragmented. We thus distinguish between the following sub-cases:
    \begin{enumerate}
        \item[a.] {When $\sigma^2/m^2 \to 0$:  The system undergoes a phase transition between homogeneous and fragmented states.} \\ 
        In this case, the outlier eigenvalue is $\lambda_{0} =  1 + m(1+\frac{\sigma^2}{m^2}) \frac{\rho_0}{\varphi} \sim 1 + m \frac{\rho_0}{\varphi}$, which is $\cO(1)$. On the contrary, the Wigner semicircle degenerates into a $\delta$-distribution, so that the outlier is always the smallest eigenvalue. Consequently, a transition from homogeneous to gregarious phase is observed when $m<-\frac{\varphi}{\rho_0}$. The heterogeneity plays no role in the phase transition.
        \item[b.] {When $\sigma^2/m^2 \to \infty$: The system is fragmented}. \\
         In this case, the spectrum is composed of the Wigner semicircle only, whose left edge diverges with $\sigma$. The associated eigenmode being characterized by $F=1$, the system is always in the fragmented phase.
        \item[c.] {When $\sigma^2/m^2 \sim \cO(1)$: The system may be homogeneous, gregarious or fragmented.} \\ 
        This is the marginal case studied in the main text, where the transitions between homogeneous, gregarious, and fragmented phases occur at finite values of $m,\sigma^2$.
    \end{enumerate}
\end{enumerate}
We highlight that case 2.c is the sole scaling for which all phase transitions occur, which motivated the study of this case in the main text. 

\section{Captions for Supplementary Movies}\label{SMsec:movies}
\begin{itemize}
    \item \textbf {Caption for Movie 1.} 
    Ecosystem fragmentation in the absence of systematic inhibition ($\sigma^2=2, m=0$) for asymmetric interactions. The simulation corresponds to panel~$\pentagon$ of Fig.~\ref{fig:snapshots}. Different colors represent different strains. Only 15 strains out of 50 are shown for better visualization.
    \item \textbf {Caption for Movie 2.} 
    Gregarious ecosystem in the absence of heterogeneity ($\sigma^2=0, m=-1$) for asymmetric interactions. The simulation corresponds to panel~{\scriptsize \faStar[regular]} of Fig.~\ref{fig:snapshots}. Different colors represent different strains. Only 15 strains out of 50 are shown for better visualization.
    \item \textbf {Caption for Movie 3.} 
    Ecosystem fragmentation in the presence of systematic inhibition ($\sigma^2=2, m=-1$) for asymmetric interactions. The simulation corresponds to panel~$\spadesuit$ of Fig.~\ref{fig:snapshots}. Different colors represent different strains. Only 15 strains out of 50 are shown for better visualization.
    \item \textbf {Caption for Movie 4.} 
    Dynamic pattern in the absence of systematic inhibition ($\sigma^2=2, m=0$) for asymmetric interactions. Different colors represent different strains. Only 15 strains out of 50 are shown for better visualization.
    \item \textbf {Caption for Movie 5.} 
    Dynamic pattern in the presence of systematic inhibition ($\sigma^2=2, m=-1$) for asymmetric interactions. Different colors represent different strains. Only 15 strains out of 50 are shown for better visualization.
\end{itemize}

\end{document}